\documentclass[a4paper,11pt]{article}
\usepackage{jcappub} 
\makeatletter
\gdef\@fpheader{ }
\makeatother

\hypersetup{
	colorlinks   = true, 
	urlcolor     = blue, 
	linkcolor    = blue, 
	citecolor    = red   
}
\renewcommand{\b}[1]{{\bf #1}}
\newcommand{\sinc}{\operatorname{sinc}}
\renewcommand{\Re}{\operatorname{Re}}
\renewcommand{\Im}{\operatorname{Im}}
\renewcommand{\boldsymbol}{\mathbf}

\title{Deviations from Gaussian White Noise in Stochastic Inflation}

\author{Zahra Ahmadi,}
\emailAdd{zr.ahmadi@ut.ac.ir}

\author{Mahdiyar Noorbala}
\emailAdd{mnoorbala@ut.ac.ir}

\affiliation{Department of Physics, University of Tehran, Iran. P.O.\ Box 14395-547}

\abstract{Stochastic inflation is widely used as a framework to study scalar field perturbations on an inflationary spacetime in a classical manner.  In Starobinsky's seminal work and most of the subsequent literature, stochastic inflation is driven by a white noise.  This is a consequence of a number of assumptions about the background metric, the window function, and the initial state.  Given that noise is the central object in this approach, it is worthwhile to investigate how the noise is modified upon relaxing some of these assumptions.  We show that while deviation from an exact de~Sitter background maintains the white character of the noise (only with a time-dependent amplitude), deviation from the Heaviside window function or the Bunch-Davies initial state can produce colored noise.  We calculate the power spectrum and the memory of the noise for a toy model with a piecewise linear window function.  We also show that, in order to produce a colored noise, the deviation from the Bunch-Davies vacuum should essentially be a sum of two-particle states.  The resulting noise is non-stationary and we find its instantaneous power spectrum in a concrete example.  Furthermore, while deviations from de~Sitter background and sharp cutoff do not affect Gaussianity, changing the initial state yields a non-Gaussian noise.}

\begin{document}

\maketitle


\section{Introduction}\label{sec:intro}

Inflation is currently the most widely accepted model for describing the early moments of the universe. This epoch is usually driven by a scalar field known as the inflaton, which is responsible for creating an accelerating expansion era~\cite{Starobinsky:1980te, Guth:1980zm, Linde:1981mu, Albrecht:1982wi}. According to the uncertainty principle, the field undergoes quantum fluctuations in the vacuum state and then because of rapid expansion of the universe they get stretched to superhorizon scales~\cite{Mukhanov:1981xt, Hawking:1982cz, Starobinsky:1982ee, Guth:1982ec, Bardeen:1983qw}. Eventually these modes exceed the Hubble radius, $H^{-1}$,  freeze and turn into classical perturbations. Later the perturbations reenter the horizon and seed the large scale structures we see today.  Essentially the same picture is valid for any other field on the inflationary background, with varying degrees of perturbation amplitude.

Despite the quantum nature of fluctuations, one can exploit the quantum-to-classical transition of the modes to formulate a classical stochastic approach for studying perturbations~\cite{Vilenkin:1983xp, Vilenkin:1983xq, Linde:1986fd, Starobinsky:1986fx, Rey:1986zk, Aryal:1987vn}.  Stochastic inflation is essentially an effective theory focusing on the dynamics of the superhorizon (IR) modes while integrating out the subhorizon (UV) modes.  As the universe expands, the UV modes that cross the horizon become classical and merge with the IR modes to form the coarse-grained field.  In fact, the subhorizon quantum modes play the role of noise in the stochastic inflation formalism.  The long-wavelength superhorizon modes are typically described by a Langevin equation, and their statistical properties --- such as their probability distribution function --- are derived from the corresponding Fokker-Planck equation~\cite{Nambu:1987ef, Nambu:1988je, Kandrup:1988sc, Nakao:1988yi, Nambu:1989uf, Mollerach:1990zf, Spokoiny:1993uc, Linde:1993xx, Linde:1996hg, Starobinsky:1994bd, Kunze:2006tu, Prokopec:2007ak, Prokopec:2008gw, Tsamis:2005hd, Enqvist:2008kt, Finelli:2008zg, Finelli:2010sh, Garbrecht:2013coa, Garbrecht:2014dca, Burgess:2014eoa, Burgess:2015ajz, Boyanovsky:2015tba, Boyanovsky:2015jen, Fujita:2017lfu, Glavan:2017jye, Gorbenko:2019rza, Mirbabayi:2019qtx, Mirbabayi:2020vyt, Cohen:2021fzf, Cruces:2021iwq, Cruces:2022imf, Palma:2023idj, Launay:2024qsm, Honda:2024evc, Cruces:2024pni, Tomberg:2024evi, Woodard:2025cez, Briaud:2025ayt, Launay:2025lnc}.  There is also a stochastic version of the $\delta N$ formalism, which combines the power of the stochastic inflation and the $\delta N$ formalism, and is extensively used in the literature~\cite{Fujita:2013cna, Fujita:2014tja, Vennin:2015hra, Vennin:2016wnk, Assadullahi:2016gkk, Grain:2017dqa, Pattison:2017mbe, Pattison:2018bct, Noorbala:2018zlv, Firouzjahi:2018vet, Biagetti:2018pjj, Noorbala:2019kdd, Pattison:2019hef, Talebian:2019opf, Ezquiaga:2019ftu, Firouzjahi:2020jrj, Ando:2020fjm, Ballesteros:2020sre, Pattison:2021oen, Figueroa:2020jkf, Figueroa:2021zah, Ahmadi:2022lsm, Animali:2022otk, Talebian:2022jkb, Nassiri-Rad:2022azj, Asadi:2023flu, Tomberg:2023kli, Mishra:2023lhe, Jackson:2024aoo, Nassiri-Rad:2025dsa, Blachier:2025iwk}.    
  
In its conventional form, stochastic inflation assumes a slowly rolling scalar field in a de~Sitter (dS) background which is initially in the Bunch-Davies vacuum state, utilizing a sharp cutoff to split it to long and short modes. As a consequence of these assumptions, the resulting noise is Gaussian and white.  The lack of memory in a white noise leads to a coarse-grained field that has Markovian dynamics.  These features simplify the formalism and the analytical calculations and provide an intuitive insight into the stochastic process under study.  Nonetheless, changing any of the aforementioned assumptions leads to deviation from Gaussian white noise.  There has been studies in the literature on the non-whiteness of the noise in stochastic inflation.  In particular, Refs.~\cite{Winitzki:1999ve, Matarrese:2003ye, Liguori:2004fa, Mahbub:2022osb} have considered the effect of switching from a sharp cutoff to a smooth window function; however, the spectral properties of the noise were not studied specifically.  There are also numerous papers on deviation from the Bunch-Davies initial state in inflation, but we are not aware of any studies in the literature on stochastic inflation.  Therefore, the need for a comprehensive investigation seems to be plausible.

The aim of this paper is to systematically study deviations from Gaussian white noise by violating the conventional assumptions above in the case of a free scalar test field.  In particular, we will consider non-dS background, non-sharp cutoff, and non-Bunch-Davies state, and see how the noise deviates in each case from a Gaussian white one.  This deviation manifests in the color profile of the noise.  We read this information through the correlation function of the noise, or its Fourier transform, i.e., the power spectrum.  In the simplest case, the amplitude of the power spectrum becomes time-dependent, while retaining its whiteness.  A more complicated case is when the power spectrum becomes frequency-dependent and the noise is no longer white.  An even more complicated situation is when the noise is non-stationary and the power spectrum becomes time-dependent, leading to the notion of instantaneous power spectrum.  The mathematical definitions and subtleties regarding the noise and its properties are summarized in appendix~\ref{app:noise} and we will use them extensively throughout the paper.  It is in terms of these properties that we quantify the deviations from Gaussian white noise; so it will pay off to skim through appendix~\ref{app:noise} beforehand, if the reader is not familiar with these ideas.

\textit{Notations and conventions:}  (\textit{i}) We use the $(-,+,+,+)$ signature for the metric and work in units in which $\hbar=c=8\pi G_N=1$.  (\textit{ii}) Throughout this paper, we shall use proper time $t$, conformal time $\tau$, and $e$-folding time $N$, interchangeably to indicate time.  As usual, dot denotes $\frac{d}{dt}$ and $' = \frac{d}{d\tau} = a\frac{d}{dt}$.  (\textit{iii}) The phrase ``power spectrum'' or simply ``power'' is employed to refer to two distinct quantities: the noise power spectrum, which we denote by $P_{\xi_\phi}$, and the dimensionless power spectrum of the field perturbations, which we denote by ${\cal P}_\phi$; see also footnote~\ref{ft:PvsP}.  Of course, the former is the central concept in this paper.  (\textit{iv}) We refer to the step function as ``sharp cutoff'', rather than ``sharp window function''.  Any window function that is not equal to the step function is called a ``non-sharp cutoff'', even if the graph of that window function itself has sharp corners as in figure~\ref{fig:window-function}.  So non-sharp cutoffs do not necessarily mean \textit{smooth} window functions.

The rest of the paper is organized as follows:  Section~\ref{sec:review} is a general review of the formalism of stochastic inflation while some commonly used formulas are deferred to appendix~\ref{app:correlators}.  In section~\ref{sec:white} we review the conventional case of white noise, with both constant and variable amplitude.  We study the effect of deviation from the sharp cutoff in section~\ref{sec:window} and the effect of deviation from the standard Bunch-Davies initial state in section~\ref{sec:state}.  Finally, we summarize and conclude in section~\ref{sec:summary}. 



\section{Review of Stochastic Inflation and the General From of Noise}\label{sec:review}

We begin with a brief review of the general formalism of stochastic inflation.  We consider a free scalar test field with the minimally coupled action
\begin{equation}
S = \int d^4x \sqrt{-g} \left[ -\frac{1}{2} g^{\mu\nu} \partial_\mu \phi \partial_\nu \phi - \frac12 m^2 \phi^2 \right],
\end{equation}
on a fixed inflationary background with the FLRW metric $-dt^2+a(t)^2d\b x\cdot d\b x$.  The Fourier mode $\hat{\phi}_{\boldsymbol{k}}(t)$ of the quantum field operator in the Heisenberg picture $\hat\phi(t,\b x)$ has an expansion in terms of creation and annihilation operators in the familiar form
\begin{equation}\label{mode-fxn}
\hat{\phi}_{\boldsymbol{k}}(t) = \int \frac{d^3x}{(2\pi)^{3/2}} \hat\phi(t,\b x) e^{-i\b k \cdot \b x} = \phi_{k}(t) \hat{a}_{\boldsymbol{k}}+\phi^*_{k}(t) \hat{a}^{\dagger}_{-\boldsymbol{k}},
\end{equation}
where $\phi_k$ is the usual mode function satisfying
\begin{equation}\label{MS}
u_k'' + \omega_k^2 u_k = 0, \qquad \text{with} \qquad \omega_k = \sqrt{k^2 + m^2a^2 - \frac{a''}{a}},
\end{equation}
in which $u_k=a\phi_k$ is the Mukhanov-Sasaki variable.

The compatibility of the equal-time canonical commutation relation $[\hat\phi(t,\b x),  \dot{\hat\phi}(t,\b x')] = i\delta(\b x-\b x')/a^3(t)$ and the conventional normalization $[\hat a_{\b k}, \hat a^\dag_{\b k'}] = \delta(\b k - \b k')$ of the creation and annihilation operators dictates the Wronskian condition
\begin{equation}\label{Wronskian}
u_k u'^*_k - u_k^* u'_k = i.
\end{equation}
Additionally, we demand that at some initial time $t_0$ (to be chosen to be past infinity in the sequel), the Hamiltonian that generates the evolution in conformal time, takes the standard form $a^3(t_0) \int d^3k \omega_k(t_0) \hat a^\dag_{\b k} \hat a_{\b k}$ (without any $\hat a \hat a$ or $\hat a^\dag \hat a^\dag$ terms), so that $\hat a^\dag_{\b k}$ indeed creates a particle with proper energy $\omega_k(t_0)/a(t_0)$ at $t_0$, and $\hat a_{\b k}$ indeed annihilates the ground state $|0\rangle$ at $t_0$.  This, together with the Wronskian condition~\eqref{Wronskian}, fixes (up to an irrelevant phase factor) the initial conditions of the mode function to be $u_k(t_0) = 1/\sqrt{2\omega_k(t_0)}$ and $u'_k(t_0) = -i\omega_k(t_0) u_k(t_0)$.  It corresponds to choosing the solution to eq.~\eqref{MS} that has the asymptotic behavior $u_k \to e^{-ik\tau}/\sqrt{2k}$ as $\tau\to-\infty$.

We emphasize that the choice of the mode function $u_k$ is not equivalent to choosing the vacuum state.  All that we have done so far is to ensure that the ket $|0\rangle$ means the vacuum in the asymptotic past.  We still have the freedom of studying a field in an arbitrary state $|\Psi\rangle$ other than $|0\rangle$.  Although $|\Psi\rangle=|0\rangle$ is the conventional choice, we will have occasion to select other states $|\Psi\rangle\neq|0\rangle$, as we will see in section~\ref{sec:state}.

The main idea of stochastic inflation is separating the Fourier modes of the field by means of a window function $W(\kappa)$ into the long mode $\hat\phi_l$ and the short mode $\hat\phi_s$ as follows:\footnote{These equations are consistent with $\hat\phi_s+\hat\phi_l=\hat\phi$ through eq.~\eqref{mode-fxn}.  This is possible because we are considering a free field for which eq.~\eqref{mode-fxn} holds; otherwise eq.~\eqref{phi-l} would have to be replaced by $\hat\phi_l:=\hat\phi-\hat\phi_s$.}
\begin{align}
\hat{\phi}_{s}(t,\b x)&=\int \dfrac{d^3k}{(2\pi)^{3/2}} W(\kappa) \hat{\phi}_{\boldsymbol{k}}(t) e^{i\boldsymbol{k}\cdot\boldsymbol{x}}, \label{phi-s} \\
\hat{\phi}_{l}(t,\b x)&=\int \dfrac{d^3k}{(2\pi)^{3/2}} \left[1-W(\kappa)\right] \hat{\phi}_{\boldsymbol{k}}(t) e^{i\boldsymbol{k}\cdot\boldsymbol{x}}. \label{phi-l}
\end{align}
Here $\kappa=k/k_{\sigma}(t)$ and $k_{\sigma}(t)=\sigma a(t) H(t)$ is a momentum cutoff specified such that $k_{\sigma}\ll a H$, for reasons to be explained later.  The window function $W(\kappa)$ is designed to quickly approach unity for $\kappa>1$ and quickly decay to zero for $\kappa<1$.  Therefore, $\hat\phi_s$ is made up of predominantly short wavelengths, while the smoothed field $\hat\phi_l$ is composed chiefly of long wavelengths.  The choice originally made for the window function by Starobinsky was the Heaviside step function, namely, $W(\kappa)=\theta(\kappa-1)$, corresponding to a sharp momentum cutoff; but we want to keep it general for now.  

We can similarly consider the velocity filed $\hat v = d\hat\phi/dN = \dot{\hat\phi}/H$, which is conveniently defined as the derivative with respect to the $e$-folding time $N$, and split it into the long mode $\hat v_l$ and the short mode $\hat v_s$ as follows:
\begin{align}
\hat v_s(t,\b x) &= \int \frac{d^3k}{(2\pi)^{3/2}} W(\kappa) \frac{d\hat\phi_{\bf k}}{dN} e^{i \b k \cdot \b x}, \\
\hat v_l(t,\b x) &= \int \frac{d^3k}{(2\pi)^{3/2}} \left[ 1 - W(\kappa) \right] \frac{d\hat\phi_{\bf k}}{dN} e^{i \b k \cdot \b x}.
\end{align}
Note that $\hat v_l$ is the long mode component of the derivative of $\hat\phi$, and it differs from the derivative of the long mode component of $\hat\phi$.  The latter is $d\hat\phi_l/dN$ and they are related by the equations of motion for the smoothed fields,
\begin{align}
\frac{d\hat\phi_l}{dN} &= \hat v_l + \hat\xi_\phi, \label{Langevin-phi-op}\\
\frac{d\hat v_l}{dN} &= -(3-\epsilon) \hat v_l - \frac{1}{H^2} (m^2 - \frac1{a^2}\nabla^2) \hat\phi_l + \hat\xi_v, \label{Langevin-v-op}
\end{align}
where the noise operators $\hat\xi_\phi$ and $\hat\xi_v$ are defined by
\begin{align}
\hat\xi_\phi(t,\b x) &= \bar\epsilon(t) \int \frac{d^3k}{(2\pi)^{3/2}} \kappa W'(\kappa) \hat\phi_{\bf k} e^{i \b k \cdot \b x}, \label{xi-phi} \\
\hat\xi_v(t,\b x) &= \bar\epsilon(t) \int \frac{d^3k}{(2\pi)^{3/2}} \kappa W'(\kappa) \frac{d\hat\phi_{\bf k}}{dN} e^{i \b k \cdot \b x}. \label{xi-v}
\end{align}
Here $W'$ means $dW/d\kappa$ and $\bar\epsilon$ is a shorthand for the frequently appearing quantity $1-\epsilon$.  Some of the important properties of the noise correlators that we will frequently make use of in the sequel are summarized in appendix~\ref{app:correlators}.

Eqs.~\eqref{Langevin-phi-op} and \eqref{Langevin-v-op} reduce to a pair of Langevin equations, when we take the following steps: First we need to choose a state $|\Psi\rangle$ to evaluate the expectation values of the smoothed fields.  Then we can see that under fairly general conditions, the commutators of the noises become irrelevant in the limit $\sigma\to0$ (c.f.\ eqs.~\eqref{[,]-sharp} and \eqref{[phi,v]-sharp}), and thus they have the behavior of classical stochastic variables (which we denote $\phi_l$, $v_l$ and $\xi_{\phi,v}$), rather than the quantum operators ($\hat\phi_l$, $\hat v_l$ and $\hat \xi_{\phi,v}$).  We can also read the two-point correlation function of the noise in the state $|\Psi\rangle$ and obtain the statistical properties of the corresponding classical stochastic variables.  We do this for a single patch, discarding the dependence on $\b x$.  The final form of the Langevin equations will be:
\begin{align}
\frac{d\phi_l}{dN} &= v_l + \xi_\phi, \label{Langevin-phi}\\
\frac{dv_l}{dN} &= -(3-\epsilon) v_l - \frac{m^2}{H^2} \phi_l + \xi_v. \label{Langevin-v}
\end{align}
We do not intend to solve this system of stochastic differential equations here, but in order to study it one first needs to know the statistical properties of the noise.  In the simplest case that was originally studied in ref.~\cite{Starobinsky:1986fx}, we have a white noise, i.e., $\langle \xi(t_1) \xi(t_2) \rangle \propto \delta(t_1-t_2)$ (c.f.\ eqs.~\eqref{[phi,phi]+0sharp}--\eqref{[v,v]+0sharp}).  We will review it in the next section, and in the sequel we consider several situations and see how the statistics of the noise deviates from that of the white noise.  For a brief review of the mathematical properties of the noise, including the relevant definitions for the white noise, see appendix~\ref{app:noise}.    


\section{White Noise with Constant and Variable Amplitude}\label{sec:white}

In this section we briefly review two cases where white noise statistics shows up.  In both cases the window function is $W(\kappa)=\theta(\kappa-1)$ and the initial state is $|0\rangle$.  In the first case the background spacetime is the exact dS space, whereas in the other case we consider an inflationary spacetime that is not exactly dS.

\subsection{White Noise with Constant Amplitude: Free Field on dS}\label{ssec:white-const}

In dS space with Hubble constant $H$, we have $a = -1/H\tau$ and the mode function is given by
\begin{equation}\label{dS-u}
u_k = \frac{1}{2} e^{i(2\nu+1)\pi/4} \sqrt{-\pi\tau}H_{\nu}(-k\tau),
\end{equation}
where $H_{\nu}$ is the Hankel function of the first kind and order $\nu = \sqrt{\frac{9}{4}-\frac{m^2}{H^2}}$, and the irrelevant overall phase can be discarded.  This mode function gives the power spectrum ${\cal P}_\phi(k,\tau)$ and, when plugged in the general expressions~\eqref{[phi,phi]+0sharp}--\eqref{[v,v]+0sharp} of appendix~\ref{app:correlators} and under the usual assumption~\eqref{sigma-interval}, yields eqs.~\eqref{[phi,phi]+0sharpdS}--\eqref{[v,v]+0sharpdS} with $\Delta \b x=0$.  This has a simple classical stochastic interpretation: The two noises $\xi_\phi$ and $\xi_v$ are given by
\begin{equation}\label{xi-vs-xi_n}
\xi_\phi = \frac{H}{2\pi} \xi_n, \qquad \xi_v = - \frac{m^2}{3H^2} \frac{H}{2\pi} \xi_n,
\end{equation}
where $\xi_n$ is a Gaussian standard white noise that satisfies
\begin{equation}
\langle \xi_n(N_1) \xi_n(N_2) \rangle = \delta(N_1-N_2).
\end{equation}
Since this is a free theory, it is not hard to expect that the application of the Wick theorem leads to Gaussian statistics.  This is performed in appendix~\ref{app:correlators} by computing higher order correlators.  

In summary, we have two noises with power spectra $P_{\xi_\phi} = \left( \frac{H}{2\pi} \right)^2$ and $P_{\xi_v} = \left (\frac{m^2}{3H^2} \frac{H}{2\pi} \right)^2$ that are constant both in time and frequency, i.e., two stationary white noises.  Of course, if a time variable other than $N$ is used, the amplitudes will depend on time, but we shall stick to $N$ (yet continue to denote the frequency conjugate to $N$ by $\omega$).

\subsection{White Noise with Variable Amplitude: Free Massless Field on Quasi-dS Space}\label{ssec:white-var}

The next case is again a free field in the vacuum $|0\rangle$ with the same window function $W(\kappa)=\theta(\kappa-1)$, but on a quasi-dS background instead of an exact dS.  For simplicity we set the mass equal to zero.  A naive comparison with the results of the previous subsection suggests that instead of the white noise $\xi_\chi$ with constant amplitude $H_0/2\pi$, we may now have a variable amplitude $H(N)/2\pi$ due to the running Hubble parameter $H(N)$.

This expectation is qualitatively correct, as the background geometry is no longer static.  We can have a more quantitative analysis by using eq.~\eqref{[phi,phi]+0sharp} for the noise amplitude.  Since we need to evaluate ${\cal P}_\phi$ at the superhorizon wavelength $k_\sigma(N)$, if we assume slow-roll was underway up until well after $N_{k_\sigma(N)}$,\footnote{We use the common notation $N_k$ for the time of the horizon crossing of the mode $k$, i.e., the $N_*$ that satisfies $k = a(N_*) H(N_*)$.  Therefore, $N_{k_\sigma(N)}$, the time of the horizon crossing of $k_\sigma(N)$, is equal to the $N_*$ that satisfies $\sigma a(N) H(N) = a(N_*) H(N_*)$.  Clearly, since $\sigma\ll1$, $N_{k_\sigma(N)}$ must be much earlier than $N$.} we can use the standard result for the power spectrum of the massless field, namely, ${\cal P}_\phi(k,N) \to (H(N_k)/2\pi)^2$.  It then follows from eq.~\eqref{[phi,phi]+0sharp} that
\begin{equation}\label{corr-Hksigma}
\langle \xi_\phi(N_1) \xi_\phi(N_2) \rangle \approx \left( \frac{H \left( N_{k_\sigma(N_1)} \right)}{2\pi} \right)^2 \delta(N_1-N_2).
\end{equation}
This is still a white noise, because of the delta function on the RHS, but it is a non-stationary white noise, i.e., one with time-dependent amplitude (c.f.\ eq.~\eqref{non-stationary-white}; for more details on the distinction between stationary and non-stationary white noise, see appendix~\ref{app:noise}).  This time-dependent amplitude is nothing but the square root of the instantaneous power spectrum $P_{\xi_\phi}(\omega,N)$ of $\xi_\phi$, whose $\omega$-independence is also an indication of the whiteness of the noise $\xi_\phi$.  As before, the absence of interactions and the choice of the vacuum state lead to the Gaussianity of the noise, too.

As a concrete example, consider an accelerating background with a perfect fluid with time-independent equation of motion parameter $-1<w<-1/3$.  This means
\begin{equation}\label{p}
a \propto \tau^p, \qquad H \propto e^{-\epsilon N}, \qquad \text{where} \qquad p = \frac{2}{1+3w} = \frac{1}{\epsilon-1}.
\end{equation}
The mode function has the same form as in eq.~\eqref{dS-u} with $\nu=1/2-p$, which upon using eq.~\eqref{[phi,phi]+0sharp} yields
\begin{equation}\label{corr-w}
\langle \xi_\phi(N_1) \xi_\phi(N_2) \rangle = \frac{\sigma^3 H(N_1)^2}{8\pi} \left| H_{\nu}(-p\sigma) \right|^2 \delta(N_1-N_2).
\end{equation}
This example has the peculiar feature that the noise amplitude depends on $\sigma$, with $\sigma$-independence restored in the limit $w\to-1$; it also shows classical behavior even away of $\sigma\to0$~\cite{Noorbala:2024fim}.  At any rate, the time-dependence of the noise amplitude in eq.~\eqref{corr-w} is given by $H(N_1)$.\footnote{At first sight, this may seem to be at odds with $H(N_{k_\sigma(N_1)})$ of eq.~\eqref{corr-Hksigma}.  However, it is easy to check that $N_{k_\sigma(N)} = N - p\log\sigma$, which when combined with eq.~\eqref{p} reproduces the extra factor $\sigma^{3-2\nu}$ that is implicit in the $\sigma\to0$ limit of eq.~\eqref{corr-w}.  The remaining coefficients in eq.~\eqref{corr-w} also tend to unity in the slow-roll limit where eq.~\eqref{corr-Hksigma} must be valid.}  In other words, the instantaneous power spectrum of $\xi_\phi$ is
\begin{equation}
P_{\xi_\phi}(\omega,N) \propto e^{-2\epsilon N},
\end{equation}
which is once again a non-stationary white noise.

\begin{figure}[h]
	\centering
	\includegraphics[width=0.6\textwidth]{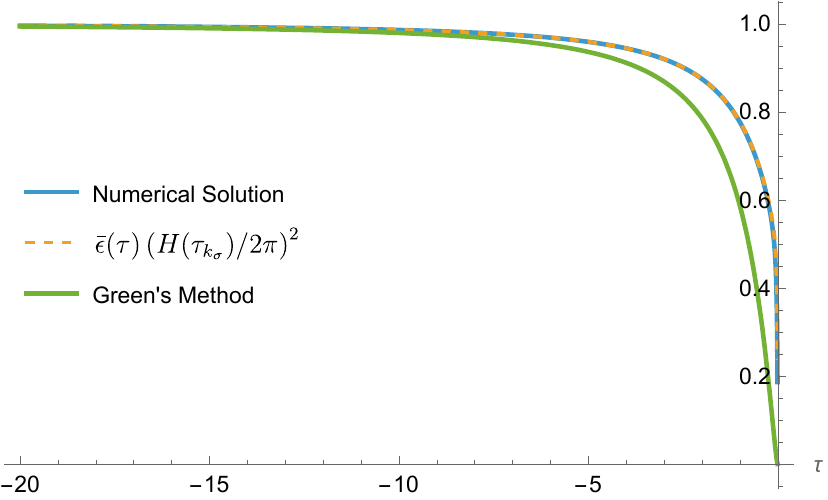}
	\caption{The power spectrum of $\xi_\phi$ in the quasi-dS toy model of eq.~\eqref{a-toy-model} (with $c=1$ and $\sigma=0.01$), calculated numerically (blue), by applying eq.~\eqref{corr-Hksigma} with $\bar\epsilon$ restored (orange), and by the analytic approximation of Green's method in appendix~\ref{app:Green} (green).  The horizontal axis has units of $H_0^{-1}$ and the vertical axis has units of $(H_0/2\pi)^2$.}
	\label{fig:variable-amplitude}
\end{figure}

We have also plotted the instantaneous power spectrum of the noise in a toy model with
\begin{equation}\label{a-toy-model}
a(\tau) = -\frac{1}{H_0\tau} + \frac{c}{H_0^2\tau^2}
\end{equation}
in figure~\ref{fig:variable-amplitude}.  The calculation is done once by numerically solving for the mode function, once using the approximation~\eqref{corr-Hksigma}, and once with an analytical approximation of the mode function that is presented in appendix~\ref{app:Green}.  Clearly, the three methods are consistent.


\section{Deviation Due to the Window Function}\label{sec:window}

\begin{figure}[h]
	\centering
	\includegraphics[width=0.6\textwidth]{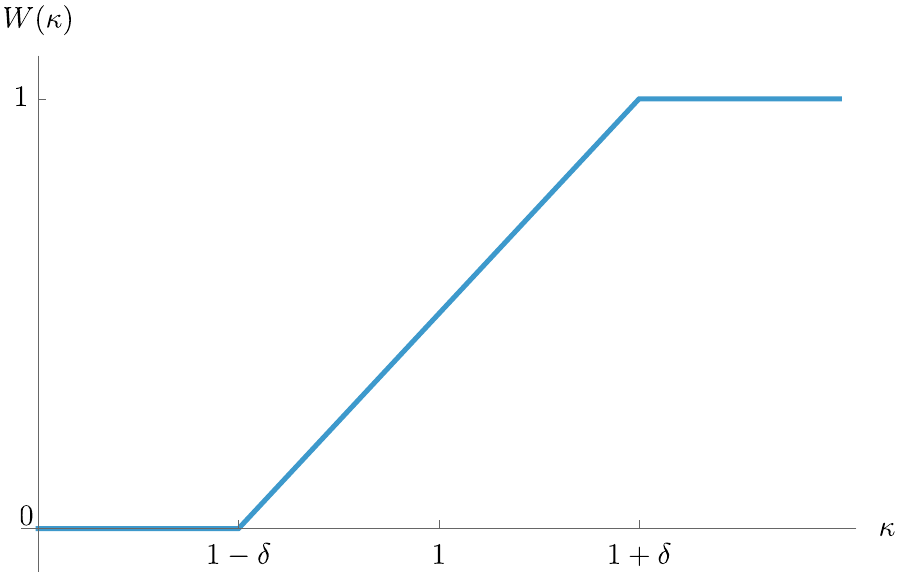}
	\caption{The window function~\eqref{W} used for our non-sharp cutoff.  The width $2\delta$ the interval of $\kappa=k/k_\sigma$ over which the modes contribute to both $\phi_l$ and $\phi_s$.}
	\label{fig:window-function}
\end{figure}

We now move on to investigate the effect of the window function on the noise statistics.  To avoid additional complications, we choose the vacuum state $|\Psi\rangle = |0\rangle$ (as we know, this guarantees Gaussianity as well).  The general expressions for this situation are given in eqs.~\eqref{[phi,phi]+0}--\eqref{[v,v]+0}.  For the purpose of illustration in the specific calculations below, we employ the piecewise linear window function
\begin{equation}\label{W}
W(\kappa) = \begin{cases}
0 & \kappa<1-\delta, \\
\dfrac12 + \dfrac{\kappa-1}{2\delta} & 1-\delta<\kappa<1+\delta, \\
1 & \kappa>1+\delta,
\end{cases}
\end{equation}
where $0<\delta<1$ represents half of the width; see figure~\ref{fig:window-function}.  This enables us to study the effect of deviation from the sharp cutoff.  Eq.~\eqref{W} implies
\begin{equation}\label{W'-delta}
W'(\kappa)= \begin{cases}
\dfrac1{2\delta} & |\kappa-1|<\delta, \\
0 & |\kappa-1|>\delta, 
\end{cases}
\end{equation}
whose $\delta\to0$ limit is the Dirac delta function $W'(\kappa)\to\delta(\kappa-1)$ that corresponds to the usual sharp cutoff.

To proceed, note that, according to eq.~\eqref{W'-delta}, the presence of the product $W'(\kappa_1)W'(\kappa_2)$ in the integrand of eq.~\eqref{[phi,phi]+0} enforces the condition
\begin{equation}\label{enforce}
\frac{|{\cal H}_2 - {\cal H}_1|}{{\cal H}_1 + {\cal H}_2} < \delta,
\end{equation}
where ${\cal H}=aH$, and the indices 1 and 2 refer to the times $N_1$ and $N_2$, respectively.  This means that the correlator is nonzero only when $-\Delta N_- < \Delta N < \Delta N_+$, where $\Delta N = N_2-N_1$, and $\Delta N_\pm$‌ (which are functions of $N_1$) are defined implicitly via\footnote{Note that according to this definition, $\Delta N_+$ and $\Delta N_-$ are related as follows.  If $N_2-N_1 = \Delta N_+(N_1)$ then $\Delta N_-(N_2) = \Delta N_+(N_1)$; and if $N_2-N_1 = -\Delta N_-(N_1)$ then $\Delta N_+(N_2) = \Delta N_-(N_1)$.  In other words, $\Delta N_-(N + \Delta N_+(N)) = \Delta N_+(N)$ and $\Delta N_+(N - \Delta N_-(N)) = \Delta N_-(N)$.  We have also used the fact that in an inflationary background, ${\cal H} = \dot a$ increases with time.}
\begin{equation}
{\cal H}(N_1\pm\Delta N_\pm) = \left[ \frac{1+\delta}{1-\delta} \right]^{\pm1} {\cal H}(N_1),
\end{equation}
which is equivalent to
\begin{equation}
\Delta N_\pm \mp \int_{N_1}^{N_1\pm\Delta N_\pm} \epsilon(N)dN = \log \frac{1+\delta}{1-\delta}.
\end{equation}
In general $\Delta N_+ \neq \Delta N_-$, but for small $\Delta N_\pm$ (compared to the time scale of variation of $\epsilon$, i.e., for $\Delta N_\pm \ll (d\log\epsilon/dN)^{-1} = \eta^{-1}$), we can approximate the integral above by $\epsilon \Delta N_\pm$ and the interval of nonvanishing correlation becomes symmetric around $N_1$:
\begin{equation}
\Delta N_+(N_1) = \Delta N_-(N_1) = \frac{1}{1-\epsilon(N_1)} \log \frac{1+\delta}{1-\delta}.
\end{equation}
Note that we have not made assumptions about the smallness of $\delta$ in this approximation, and that it becomes exact when $\epsilon$ is constant.

The condition~\eqref{enforce} implies that the correlator $\langle \xi_\phi(N_1) \xi_\phi(N_2) \rangle$ is proportional to the product $\theta(\Delta N_+ - \Delta N) \theta(\Delta N + \Delta N_-)$ of step functions, which in the case of symmetric interval becomes $\theta(\Delta N_\pm - |\Delta N|)$.  The presence of this step function indicates that the noises are uncorrelated at time separations greater than $\Delta N_\pm$ (depending on whether we look at future or past, respectively).  In other words, the coarse-grained field has memory of the past $\Delta N_-$ $e$-folds, but beyond that there is no memory.  The origin of this memory (equivalently, this correlation between noise at unequal times) lies in the choice of the window function, as follows:  Every mode $\bf k$ crosses the cutoff $k_\sigma(N)$ at a time $N$.  With a sharp cutoff $W=\theta(\kappa-1)$, as soon as this crossing happens, the mode $\bf k$ becomes part of the noise at time $N$; and remarkably, $\bf k$ won't contribute to the noise at any time other than $N$.  With any other window function, on the contrary, the mode $\b k$ contributes to the noise at multiple times other than the cutoff crossing time $N$, thereby establishing a correlation between the noise at unequal times.  Note that this argument relies on the fact that in a free field all Fourier modes are independent.

\begin{figure}
	\centering
	\begin{minipage}{0.49\textwidth}
		\centering
		\includegraphics[width=\textwidth]{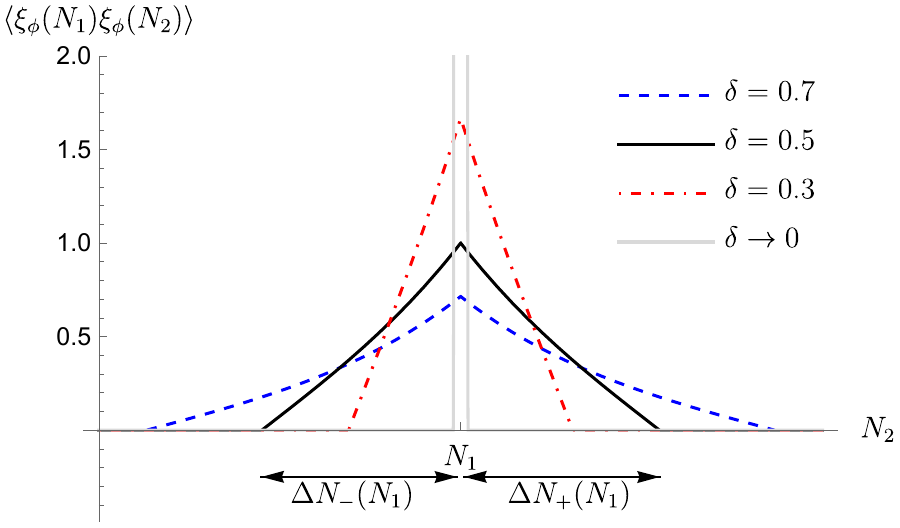}
	\end{minipage}
	\begin{minipage}{0.49\textwidth}
		\centering
		\includegraphics[width=\textwidth]{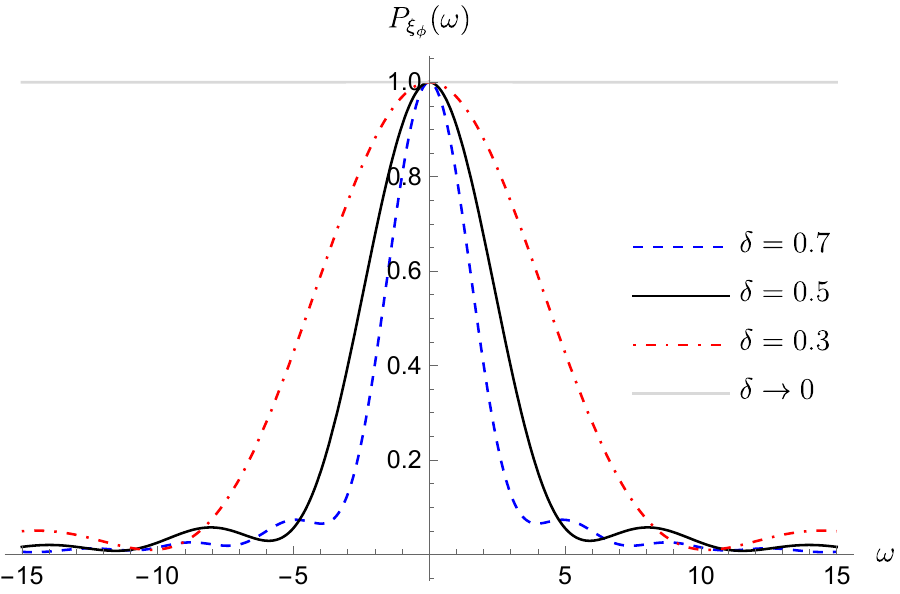}
	\end{minipage}
	\caption{The noise statistics for a massless field in exact dS, using the window function~\eqref{W}.  Left: The correlator~\eqref{corr-window-dS}.  Right: The power spectrum~\eqref{power-window-dS}.  In both cases the vertical axis has units of $(H/2\pi)^2$.}
	\label{fig:noise-window}
\end{figure}

As a concrete example, we can work out the case of a massless field on exact dS background.  We have
\begin{equation}\label{DN-vs-delta}
\Delta N_\pm = \log \frac{1+\delta}{1-\delta}, \qquad \text{i.e.,} \qquad \delta = \tanh \frac{\Delta N_\pm}{2};
\end{equation}
and the mode function is famously given by
\begin{equation}\label{u-dS-m=0}
u_k(\tau) = \frac{1}{\sqrt{2k}} \left( 1 - \frac{i}{k\tau} \right) e^{-ik\tau}.
\end{equation}
The effect of the $\kappa W'$ factors in the integral~\eqref{[phi,phi]+0} for $\langle \xi_\phi(\tau_1) \xi_\phi(\tau_2) \rangle$ is to set the integration domain to be (for $\tau_2>\tau_1$):
\begin{equation}
k \in [(1-\delta)k_\sigma(\tau_2), (1+\delta)k_\sigma(\tau_1)] = [-(1-\delta) \sigma/\tau_2, -(1+\delta) \sigma/\tau_1].
\end{equation}
Performing the integral and taking the $\sigma\to0$ limit, we obtain (still for $\tau_2>\tau_1$):
\begin{equation}
\langle \xi_\phi(\tau_1) \xi_\phi(\tau_2) \rangle = \frac1{8\delta^2} \left[ (1+\delta)^2 \frac{\tau_2}{\tau_1} - (1-\delta)^2 \frac{\tau_1}{\tau_2} \right] \left( \frac{H}{2\pi} \right)^2 \theta \left( \log \frac{1+\delta}{1-\delta} - \Delta N \right).
\end{equation}
In terms of the number of $e$-folds, we have
\begin{equation}\label{corr-window-dS}
\langle \xi_\phi(N_1) \xi_\phi(N_2) \rangle = \frac{\sinh (\Delta N_\pm - |\Delta N|)}{4\sinh^2 (\Delta N_\pm/2)} \left( \frac{H}{2\pi} \right)^2 \theta ( \Delta N_\pm - |\Delta N| ),
\end{equation}
which is now valid for all values of $\Delta N$ and is plotted in the left panel of figure~\ref{fig:noise-window}.  We can clearly see that the noise memory is $\Delta N_\pm$ $e$-folds.  Also as a confirmation, it is straightforward to check that in the $\Delta N_\pm\to0$ limit (corresponding to $\delta\to0$), we recover the conventional result $\langle \xi_\phi(N_1) \xi_\phi(N_2) \rangle = (H/2\pi)^2 \delta(N_1-N_2)$, i.e., a white noise without memory.

The correlator~\eqref{corr-window-dS} clearly depends only on the difference $\Delta N$ of times, so it is a stationary process.  We can thus compute the power spectrum of the noise by taking the Fourier transform of the correlator, yielding
\begin{equation}\label{power-window-dS}
P_{\xi_\phi}(\omega) = \left( \frac{H}{2\pi} \right)^2 \frac{\cosh \Delta N_\pm - \cos (\omega \Delta N_\pm)}{\cosh \Delta N_\pm - 1} \frac{1}{\omega^2+1}.
\end{equation}
Evidently, this is not a white noise since $P_{\xi_\phi}(\omega)$ is not constant. Thus we have a (Gaussian) colored noise.  The color profile is depicted in the right panel of figure~\ref{fig:noise-window}, where high frequency components are seen to decay in an oscillatory manner.  Of course, in the limit $\Delta N_\pm\to0$, we recover the $\omega$-independent white noise $P_{\xi_\phi}(\omega) = (H/2\pi)^2$.

The opposite extreme is $\delta\to1$, corresponding to $\Delta N_\pm\to\infty$, i.e., infinite memory.  In this limit, we find
\begin{equation}\label{corr-1/a}
\langle \xi_\phi(N_1) \xi_\phi(N_2) \rangle = \frac12 \left( \frac{H}{2\pi} \right)^2 e^{-|\Delta N|},
\end{equation}
and
\begin{equation}
P_{\xi_\phi}(\omega) = \left( \frac{H}{2\pi} \right)^2 \frac{1}{\omega^2+1}.
\end{equation}
The $1/a$ falloff of the correlator in eq.~\eqref{corr-1/a} is specific to our particular window function~\eqref{W} with $\delta=1$.  It is found in ref.~\cite{Winitzki:1999ve} that for a Gaussian window function,
\begin{equation}
\langle \xi_\phi(N_1) \xi_\phi(N_2) \rangle = \left( \frac{H}{2\pi} \right)^2 \frac{1}{2\cosh^2\Delta N},
\end{equation}
which yields the power spectrum
\begin{equation}
P_{\xi_\phi}(\omega) = \left( \frac{H}{2\pi} \right)^2 \frac{\pi\omega/2}{\sinh(\pi\omega/2)}.
\end{equation}
Furthermore, it is shown that the correlator decays like $1/a^2$ for a large class of smooth window functions, i.e.,
\begin{equation}
\langle \xi_\phi(N_1) \xi_\phi(N_2) \rangle \sim e^{-2|\Delta N|},
\end{equation}
which implies the power spectrum
\begin{equation}
P_{\xi_\phi}(\omega) \sim \frac{1}{(\omega/2)^2+1}.
\end{equation}
Our window function~\eqref{W}, which is chosen solely for the purpose of illustration of the effect of the width, lacks the nice smooth properties of ref.~\cite{Winitzki:1999ve} and therefore has a different behavior.  There are even smooth window functions that lack those properties, e.g., those that are considered in ref.~\cite{Mahbub:2022osb} for which the falloff behavior is like $\sim 1/a^n$.  There are also closed form expressions for the correlator when $W'(\kappa)$ has a Gaussian shape around $\kappa=1$~\cite{BahmaniNamjoo}.

Let us make a final comment.  We have studied the spectral property of the noise in a concrete example where we obtained the correlator~\eqref{corr-window-dS}.  It is conceivable that if we slightly change the background from exact dS, the correlator changes accordingly, but still retains the step functions $\theta(\Delta N_+ - \Delta N) \theta(\Delta N + \Delta N_-)$.  Naively, one may expect that by a suitable choice of the background, one can construct any correlator of the form $\langle \xi_\phi(N_1) \xi_\phi(N_2) \rangle = f(\Delta N)\theta(\Delta N_+ - \Delta N) \theta(\Delta N + \Delta N_-)$ for arbitrary functions $f$.  But that's not true.  For example, even the simple case $\langle \xi_\phi(N_1) \xi_\phi(N_2) \rangle = \theta ( \Delta N_\pm - |\Delta N| )$ does not represent a legitimate correlator, since the resulting power $P_{\xi_\phi}$ obtained by the Fourier transformation of the step function is not everywhere positive in the $\omega$-space, a property that the power must have.  Eq.~\eqref{corr-window-dS} is therefore a nontrivial correlator derived from the simple window function~\eqref{W}.


\section{Deviation Due to the Initial State}\label{sec:state}

The initial condition in inflationary models is generally considered to be Bunch-Davies. This is a quantum vacuum state that assumes that the modes deep inside the horizon are in their ground state. This assumption, together with taking a sharp cutoff for dividing long and short modes, simplifies the calculations in stochastic inflation and leads to a white noise.

We now consider deviation from the vacuum state and its effects on the noise.  It is well known that for the inflaton field, choosing a non-Bunch-Davies initial condition can cause additional interactions between modes which lead to non-Gaussianities. Since no significant amount of non-Gaussianity is expected, one usually looks for a small deviation from vacuum.  For a test field, there is in general no observational constraint, but for the sake of simplicity, we choose to work with small deviations from vacuum, regardless of the nature of the scalar field.

In the Heisenberg picture, the state in its most general form can be written as follows:
\begin{equation}\label{Psi-CN}
\left|\Psi\right\rangle=C_{0}\left|0\right\rangle+\sum_{N=1}^{\infty}\int d^{3}q_{1}\ldots\int d^{3}q_{N} \frac{1}{N!} C_{N}\left(\boldsymbol{q}_{1},\ldots,\boldsymbol{q}_{N}\right)\left| \boldsymbol{q}_{1},\ldots,\boldsymbol{q}_{N}\right\rangle,
\end{equation}
which is the linear combination of all possible arrangements of initial particles. The ket
\begin{equation}
\left| \boldsymbol{q}_{1},\ldots,\boldsymbol{q}_{N} \right\rangle = \hat{a}^{\dagger}_{\boldsymbol{q}_1} \ldots \hat{a}^{\dagger}_{\boldsymbol{q}_N} \left|0\right\rangle
\end{equation}
represents the state for an $N$-particle configuration, where $\boldsymbol{q}_{i}$ is the $i$-th particle (comoving) momentum and $C_{N}$ is the complex amplitude of finding the system in the state $\left| \boldsymbol{q}_{1},\ldots,\boldsymbol{q}_{N} \right\rangle$. Moreover, $C_{N}$ has to be symmetric under the exchange of any pair of $\boldsymbol{q}_{i}$s as $\left|\Psi\right\rangle$ represents bosonic particles.  This symmetry is also manifest in the orthogonality relation of the states, namely,
\begin{equation}\label{orthogonality}
\left\langle \boldsymbol{p}_{1},\ldots,\boldsymbol{p}_{M} |\boldsymbol{q}_{1},\ldots,\boldsymbol{q}_{N} \right\rangle = \delta_{MN} \sum_{\mathbb{P}} \prod_{i=1}^N \delta(\boldsymbol{p}_{i} - \boldsymbol{q}_{\mathbb{P}(i)}),
\end{equation}
where the sum is over all permutations $\mathbb{P}$ of $N$ objects.  We should also mention that $\left|0\right\rangle$ refers to the ground state, i.e., the Bunch-Davies vacuum, so taking $|C_{0}|\gg |C_{1}|, |C_{2}|,  \ldots$ ensures that  $\left|\Psi\right\rangle$ is a small deviation from the Bunch-Davies vacuum.

For the purposes of this paper, we limit our calculations to the next-to-leading order in the small deviations from vacuum.  Note that from the normalization condition $\langle\Psi|\Psi\rangle=1$, we have 
\begin{equation}\label{CN-normalization}
|C_0|^2 + \sum_{N=1}^{\infty}\int d^{3}q_{1}\ldots\int d^{3}q_{N} \frac{1}{N!} |C_{N}|^2 = 1.
\end{equation}
Thus if $\varepsilon\ll1$ denotes the small deviation from the Bunch-Davies vacuum, then we have
\begin{equation}
|C_0|^2 = 1 - O(\varepsilon^2); \qquad \forall N>0: |C_N| = O(\varepsilon).
\end{equation}
Therefore, to calculate the next-to-leading order (i.e., $O(\varepsilon)$) correction to observables (which are expectation values bilinear in $|\Psi\rangle$), it is sufficient to consider terms that contain products of $C_0$ and only one $C_{N>0}$.

We are interested in calculating the noise correlation function for the initial state~\eqref{Psi-CN}. In the general case the correlation function is related to the anticommutator which is given by eq.~\eqref{[phi,phi]+}, which for $\b x_1=\b x_2=0$ takes the following form: 
 \begin{multline}\label{xixi-Psi}
      \left\langle\Psi\right| \{ \hat{\xi}_\phi \left(N_{1}\right), \hat{\xi}_\phi \left(N_{2}\right) \} \left|\Psi\right\rangle=\frac{\bar\epsilon(N_{1}) \bar\epsilon(N_{2})}{\left(2\pi\right)^{3}} \int d^3k_{1}\int d^3k_{2} \\
  \frac{k_{1}}{k_{\sigma}(N_1)} W^\prime\left(\frac{k_{1}}{k_{\sigma}(N_1)}\right) \frac{k_{2}}{k_{\sigma}(N_2)} W^\prime\left(\frac{k_{2}}{k_{\sigma}(N_2)}\right)  2\Re \left\langle\Psi\right|\hat{\phi}_{\boldsymbol{k}_{1}}(N_{1}) \hat{\phi}_{\boldsymbol{k}_{2}}(N_{2})\left|\Psi\right\rangle.
 \end{multline}
Substituting $\hat{\phi}_{\boldsymbol{k}}$ from eq.~\eqref{mode-fxn} in the operator part of this expression, we find for the matrix element:
\begin{equation}\label{<aa+aa+aa+aa>}
 \langle\Psi| \phi_{k_{1}}\phi_{k_{2}} \hat{a}_{\boldsymbol{k}_{1}}\hat{a}_{\boldsymbol{k}_{2}}+\phi_{k_{1}}\phi_{k_{2}}^{*} \hat{a}_{\boldsymbol{k}_{1}}\hat{a}_{-\boldsymbol{k}_{2}}^\dagger+\phi_{k_{1}}^{*}\phi_{k_{2}} \hat{a}_{-\boldsymbol{k}_{1}}^\dagger \hat{a}_{\boldsymbol{k}_{2}}+\phi_{k_{1}}^{*}\phi_{k_{2}}^{*} \hat{a}_{-\boldsymbol{k}_{1}}^\dagger \hat{a}_{-\boldsymbol{k}_{2}}^\dagger
  |\Psi\rangle.
\end{equation}
By inspecting this relation, it is clear that the only term from $\left|\Psi\right\rangle$ that can be involved in the next-to-leading order, corresponds to the two-particle states, i.e., $C_2$. This is because the combination of $\hat a$ and $\hat a^\dagger$ operators is such that it either lowers the number of particles twice or raises them twice.\footnote{Note that the situation in which the number of particles stays unchanged, makes either a leading order term ($O(|C_0|^2) = 1 - O(\varepsilon^2)$), or a second order term ($O(C_N^*C_N) = O(\varepsilon^2)$ for $N>0$), both of which have no $O(\varepsilon)$ effect.}
So for our next-to-leading order purposes of this paper, we work with the state
\begin{equation}\label{Psi-C2}
    \left|\Psi\right\rangle=C_{0} \left|0\right\rangle + \frac12 \int d^{3}q_{1} \int d^{3}q_{2} C_{2}\left(\boldsymbol{q}_{1},\boldsymbol{q}_{2}\right) \left|\boldsymbol{q}_{1}, \boldsymbol{q}_{2} \right\rangle.
\end{equation}
Let us note, in passing, that the total comoving momentum in this state has expectation value and variance determined by
\begin{equation}\label{p-C2}
\langle \Psi | \hat{\b p} | \Psi \rangle = \frac12 \int d^3q_1 \int d^3q_2 |C_{2}(\b q_1,\b q_2)|^2 (\b q_1 + \b q_2),
\end{equation}
and
\begin{equation}\label{p2-C2}
\langle \Psi | \hat{\b p}^2 | \Psi \rangle = \frac12 \int d^{3}q_{1} \int d^{3}q_{2} |C_{2}\left(\boldsymbol{q}_{1},\boldsymbol{q}_{2}\right)|^2 (\b q_1 + \b q_2)^2.
\end{equation}
Plugging the state~\eqref{Psi-C2} in eq.~\eqref{xixi-Psi} and using the orthogonality relation~\eqref{orthogonality}, we find the next-to-leading correction to the anticommutator to be
\begin{multline}
    \left\langle\Psi\right| \{ \hat{\xi}_\phi \left(N_{1}\right), \hat{\xi}_\phi \left(N_{2}\right) \} \left|\Psi\right\rangle_{\text{NLO}} =\frac{\bar\epsilon(N_{1}) \bar\epsilon(N_{2})}{\left(2\pi\right)^{3}} \int d^3k_{1}\int d^3k_{2} \\
 \frac{k_{1}}{k_{\sigma}(N_1)} W^\prime\left(\frac{k_{1}}{k_{\sigma}(N_1)}\right) \frac{k_{2}}{k_{\sigma}(N_2)} W^\prime\left(\frac{k_{2}}{k_{\sigma}(N_2)}\right) 4 \Re\left[C_{0}^*C_{2}\left(\boldsymbol{k}_{1},\boldsymbol{k}_{2}\right)\phi_{k_{1}}(N_{1})\phi_{k_{2}}(N_{2})\right],
\label{non-BD-correlation-W}
\end{multline}
where the subscript NLO denotes the $O(\varepsilon)$-part of the expression.\footnote{Had we worked with the inflaton field, instead of a test field, we would have other contributions to the NLO result, too.  Those would arise because of the backreaction of the noise on the background spacetime, which would manifest itself through $O(\varepsilon)$-modifications of $\bar\epsilon$ and $\phi_k(N)$.  However, these modifications can only produce white noise corrections, unless a non-sharp cutoff is used (as in the previous section), because, being already of $O(\varepsilon)$, they have to be evaluated in the leading order state $|\Psi\rangle = |0\rangle$.}

To go further, we specialize to the case of the sharp cutoff, where the delta function $W'(\kappa) = \delta(\kappa-1)$ eliminates two of the six integrations in eq.~\eqref{non-BD-correlation-W}.  We also assume that $C_2$ is a function only of the magnitudes of its arguments, i.e., $C_2(\b q_1, \b q_2) = C_2(q_1,q_2)$, so that, according to eq.~\eqref{p-C2}, the total momentum of the particles vanishes and therefore there is no preferred direction in the universe (but note that this doesn't make $\langle \hat{\b p}^2 \rangle$ zero).  This assumption enables us to handle the remaining four angular integrals, and the NLO correction to the correlation function takes the simple form
\begin{equation}\label{non-BD-correlation-sharp}
    \left\langle\Psi\right| {\xi}_\phi \left(N_{1}\right) {\xi}_\phi \left(N_{2}\right)\left|\Psi\right\rangle_{\text{NLO}} =\frac4\pi \bar\epsilon(N_{1}) \bar\epsilon(N_{2}) k_1^3 k_2^3 \Re \left[ C_{0}^*C_{2}(k_1,k_2) \phi_{k_1}(N_{1}) \phi_{k_2}(N_2) \right],
\end{equation}
where $k_i = k_\sigma(N_i)$.  Of course, this expression is to be added to the $O(\varepsilon^0)$ part of the correlator, which is a white noise proportional to $\delta(N_1-N_2)$.  However, eq.~\eqref{non-BD-correlation-sharp} is clearly not proportional to $\delta(N_1-N_2)$, so the total noise is no longer white.  In fact, eq.~\eqref{non-BD-correlation-sharp} is not even a function of the time difference $N_1-N_2$, and so we do not have a stationary process to begin with.  Below, we study the spectral properties of this noise in a simple setup.  But let us note, beforehand, that in order to obtain a sizable NLO correction of this type between times $N_1$ and $N_2$, we need to arrange for $C_2(q_1,q_2)$ to be large at the corresponding modes $q_1 = k_\sigma(N_1)$ and $q_2 = k_\sigma(N_2)$.  Put in other words, if $C_2$ is large at some point $(q_1, q_2)$, then the correlator receives a significant correction when $N_1$ and $N_2$ are the horizon exit times of $q_1/\sigma$ and $q_2/\sigma$, correspondingly.

Let us consider once again the case of a massless field on an exact dS background.  Using the mode function~\eqref{u-dS-m=0} and setting $\epsilon=0$ in eq.~\eqref{non-BD-correlation-sharp}, we obtain
\begin{equation}\label{non-BD-correlation-dS}
    \left\langle\Psi\right| {\xi}_\phi \left(N_{1}\right) {\xi}_\phi \left(N_{2}\right)\left|\Psi\right\rangle_{\text{NLO}} = \frac{2}\pi \sigma^3H^5 e^{3(N_1+N_2)/2} \Re \left[ (\sigma+i)^2 e^{-2i\sigma} C_{0}^*C_{2}(\sigma H e^{N_1},\sigma H e^{N_2}) \right].
\end{equation}
As we have explained in appendix~\ref{app:noise}, such a correlator falls in the category of non-stationary processes for which the appropriate substitute for the notion of power spectrum is the instantaneous power spectrum given by eq.~\eqref{inst-power}.  Evidently, the leading order correlator yields the usual power
\begin{equation}
P_{\xi_\phi}^{\text{LO}}(\omega,N_0) = |C_0|^2 \left( \frac{H}{2\pi} \right)^2,
\end{equation}
which is independent of $N_0$ as it should.  The NLO correction, however, is not so, as we compute now.  To simplify the matters, we will also assume that $C_0$ and $C_2$ are real, and then in the small-$\sigma$ limit, we obtain the NLO correction to the (instantaneous) power from eq.~\eqref{inst-power} to be
\begin{equation}
P_{\xi_\phi}^{\text{NLO}}(\omega,N_0) = -\frac{2}\pi \sigma^3H^5 e^{3N_0} C_0 {\cal C}_2(\omega,N_0),
\end{equation}
where ${\cal C}_2$, which is essentially a Mellin transform\footnote{The Mellin transform of $f(x)$ is defined by $\int_0^\infty x^{s-1} f(x) dx$.} of $C_2$, is given by
\begin{equation}
\begin{aligned}
{\cal C}_2(\omega,N_0) &= \int_{-\infty}^\infty C_2(\sigma H e^{N_0-N/2},\sigma H e^{N_0+N/2}) e^{i\omega N} dN \\
&= 2 \int_0^\infty C_2(k_0x,k_0/x) x^{2i\omega-1} dx,
\label{def:calC}
\end{aligned}
\end{equation}
in which $k_0 = \sigma H e^{N_0}$ is the mode that crosses the cutoff at $N_0$.  We observe that the entire $\omega$-dependence of the power comes from ${\cal C}_2$, i.e., it is $C_2$ alone that determines the color profile.

\begin{figure}[h]
	\centering
	\begin{minipage}{0.49\textwidth}
		\centering
		\includegraphics[width=\textwidth]{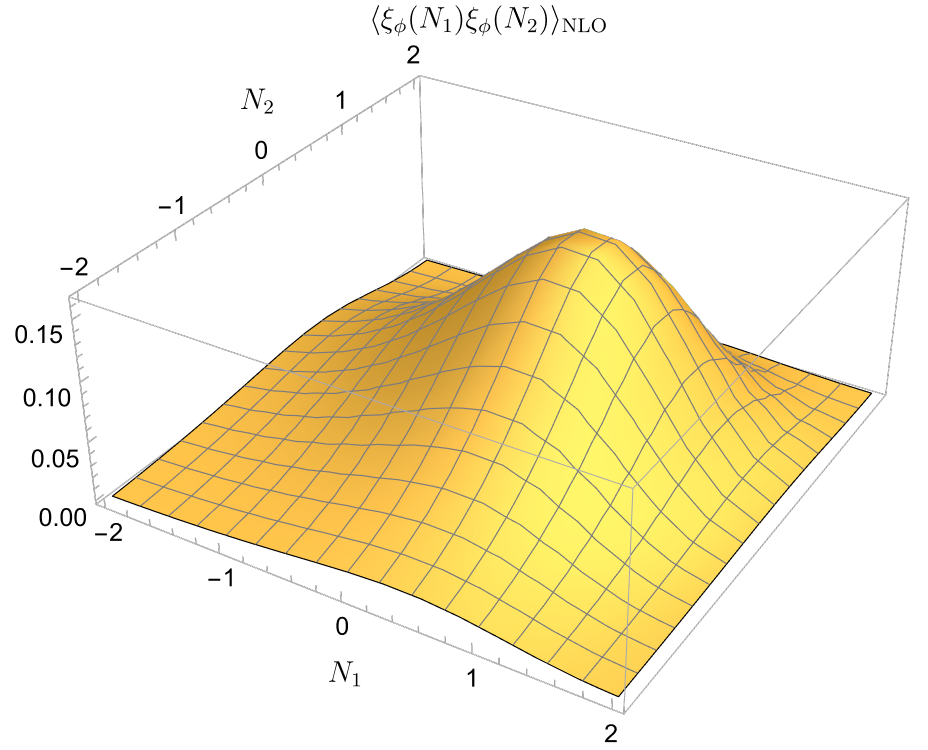}
	\end{minipage}
	\begin{minipage}{0.49\textwidth}
		\centering
		\includegraphics[width=\textwidth]{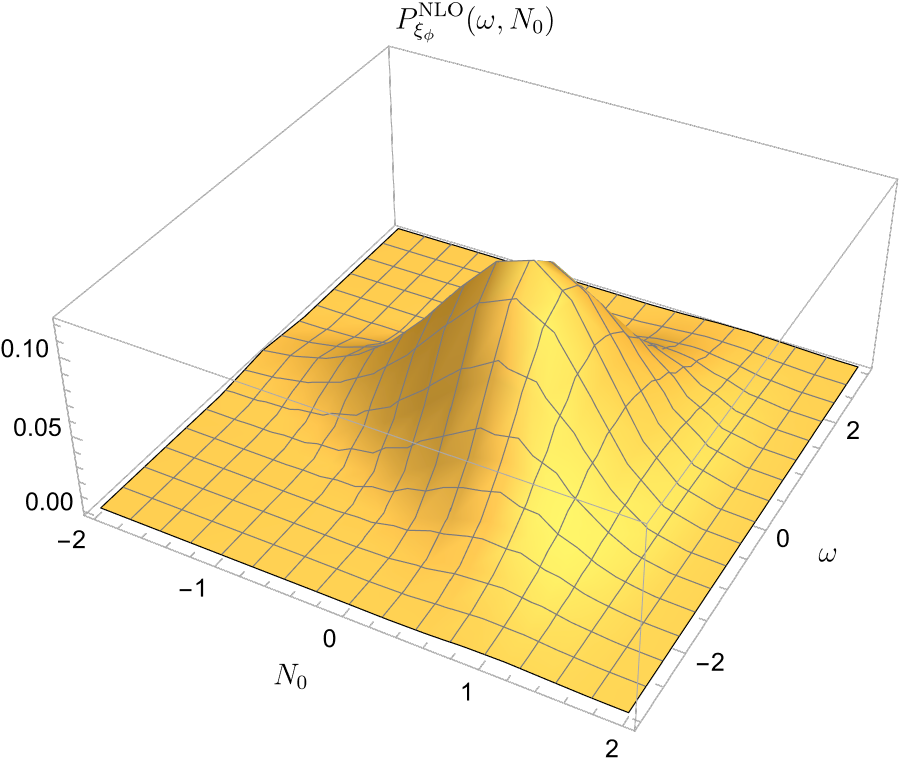}
	\end{minipage}
	\caption{The NLO part of the correlator (left) and the instantaneous power spectrum (right) for the noise of a massless field on dS in the state~\eqref{Psi-C2} with $C_2$ given by eq.~\eqref{C2-exp}.  In both cases the vertical axis has units of $-8\sqrt2\varepsilon (H/2\pi)^2$, and $N_Q$ is chosen at the origin of the time axes.}
	\label{fig:noise-state}
\end{figure}

As a concrete model, let us choose the exponentially decaying two-particle wave function
\begin{equation}\label{C2-exp}
C_2(q_1,q_2) = \frac{\sqrt2 \varepsilon}{\pi Q^3} \exp \left[ -\frac{q_1+q_2}{Q} \right],
\end{equation}
where the normalization is chosen such that $\frac12 \int d^3q_1 d^3q_2 |C_2|^2 = \varepsilon^2$ and hence $C_0^2 = 1 - \varepsilon^2$ (c.f.\ eq.~\eqref{CN-normalization}).  The decay exponent scale $Q$ is related to the variance of total momentum via $\langle \hat{\b p}^2 \rangle = 6 \varepsilon^2 Q^2$ (c.f.\ eq.~\eqref{p2-C2}).  We may also write this dimensionful parameter as $Q = k_\sigma(N_Q)$ and identify $N_Q$ as the time that $Q$ crosses the cutoff.  This makes $N_Q$ a privileged time: the modes that cross the cutoff after $N_Q$ are exponentially suppressed in the expansion~\eqref{Psi-C2} of the state $|\Psi\rangle$.  The correlator for this model can now be computed as
\begin{equation}\label{NLO-correlator-Psi}
    \left\langle\Psi\right| {\xi}_\phi \left(N_{1}\right) {\xi}_\phi \left(N_{2}\right)\left|\Psi\right\rangle_{\text{NLO}} = - 8\sqrt2 \varepsilon \left( \frac{H}{2\pi} \right)^2 \left[ a_Q(N_1) a_Q(N_2) \right]^{3/2} e^{-a_Q(N_1) - a_Q(N_2)},
\end{equation}
where $a_Q(N) = \exp(N-N_Q)$ is the expansion factor between $N_Q$ and $N$.  On the other hand, the Mellin transform~\eqref{def:calC} gives
\begin{equation}
{\cal C}_2 (\omega,N_0) = \frac{\sqrt2 \varepsilon}{\pi Q^3} K_{-2i\omega} (2k_0/Q),
\end{equation}
where $K_\nu(z)$ is the modified Bessel function of the second kind.  Putting everything together, we can finally obtain the instantaneous power spectrum of this model:
\begin{equation}\label{inst-power-Psi}
P_{\xi_\phi}(\omega,N_0) = \left( \frac{H}{2\pi} \right)^2 \left[ 1 - 8\sqrt2 \varepsilon a_Q(N_0)^3 K_{-2i\omega} \left( 2a_Q(N_0) \right) + O(\varepsilon^2) \right].
\end{equation}
We have plotted the NLO part of the correlator as well as the power in figure~\ref{fig:noise-state}.  Note how the power is suppressed/enhanced (for positive/negative $\varepsilon$) around a certain time $\sim N_Q$.  More quantitatively, the NLO correction to the correlator (eq.~\eqref{NLO-correlator-Psi}) is maximal at $a_Q(N_0)=3/2$, corresponding to $N_1 = N_2 = N_Q + \log(3/2)$.  This is about the same time that the instantaneous power~\eqref{inst-power-Psi} is maximal, namely, $a_Q(N_0)=1.268$.  To get a rough idea of the typical frequency of the non-white part of the power, we can calculate its dispersion, i.e., $\Delta\omega^2 = [\int \omega^2 {\cal C}_2 d\omega]/[\int {\cal C}_2 d\omega]$, which turns out to be $\Delta\omega^2=a_Q(N_0)/2$.  Thus the typical frequency at which the power of the noise is enhanced/suppressed is of order one (had we used $t$ instead of $N$, this typical frequency would have been of order $1/H$).  Of course, this suppression/enhancement is small, since our calculation is carried out for $\varepsilon\ll1$.

Finally, notice that although we have not calculated the higher order correlators in this section, but it should be clear that Gaussianity is in general lost.  The amount of this non-Gaussianity is evidently proportional to the deviation from the vacuum state $|0\rangle$, which we have denoted by $\varepsilon$ here.

Let us make some closing remarks in this section.  The first one is on the effect of non-Bunch-Davies states that arise from a Bogoliubov transformation.  If $|\Psi\rangle$ is such a state, it is annihilated not by $\hat a_{\b k}$, but rather by $\hat b_{\b k}$, which is a linear combination of $\hat a_{\b k}$ and $\hat a^\dag_{-\b k}$, such that
\begin{equation}
\phi_{k}(t) \hat{a}_{\boldsymbol{k}}+\phi^*_{k}(t) \hat{a}^{\dagger}_{-\boldsymbol{k}} = \varphi_{k}(t) \hat{b}_{\boldsymbol{k}}+\varphi^*_{k}(t) \hat{b}^{\dagger}_{-\boldsymbol{k}}.
\end{equation}
The function $\varphi_k(t)$ determines the relationship between the two sets of operators $\hat a_{\b k}$ and $\hat b_{\b k}$, and thus defines our Bogoliubov transformation.  It has to satisfy the same equation as $\phi_k(t)$ does, i.e., eq.~\eqref{MS}, albeit with different initial conditions.  As is well known, this state has no $b$-type particles, so we denote it by $|0_b\rangle$ (in a notation that would denote our previous vacuum by $|0_a\rangle$).  In contrast, it contains arbitrarily large number of $a$-type particles:
\begin{equation}\label{b-vs-a}
|0_b\rangle = \bigotimes_{\bf k} \sum_{n=0}^\infty A_{n,k} |n_{\bf k}, n_{-\bf k}\rangle,
\end{equation}
where $A_{n,k}$ is some coefficient that depends on the mode functions, and $|n_{\bf k}, n_{-\bf k}\rangle$ means $n$ $a$-type particles with momentum $\bf k$ and $n$ $a$-type particles with momentum $-\bf k$.  Therefore, $|0_b\rangle$ is not of the form~\eqref{Psi-C2}.  However, it is easy to see that the quantity $\langle \Psi | \hat{\phi}_{\boldsymbol{k}_{1}}(N_{1}) \hat{\phi}_{\boldsymbol{k}_{2}}(N_{2}) |\Psi \rangle$ appearing in eq.~\eqref{xixi-Psi} is again given by eq.~\eqref{<aa+aa+aa+aa>}, except that $\hat a_{\b k}$s are replaced by $\hat b_{\b k}$s, $\phi_k$s by $\varphi_k$s, and $|\Psi\rangle$ by $|0_b\rangle$.  Thus with a sharp cutoff we roll back to the old result eq.~\eqref{[phi,phi]+0sharp}, except that ${\cal P}_\phi$ is now replaced by ${\cal P}_\varphi$.  This correlator is proportional to $\delta(N_1-N_2)$ and is thus white, possibly with a variable amplitude.  Therefore, this type of state cannot produce the rich deviation from Gaussian white noise that eq.~\eqref{Psi-C2} can.

The second comment is on a sufficient condition for whiteness (in the case of sharp cutoff, but generic state).  In light of eq.~\eqref{xixi-Psi}, one such condition is that the matrix element $\langle \Psi | \hat\phi_{{\bf k}_1}(N_1) \hat\phi_{{\bf k}_2}(N_2) | \Psi \rangle$ be proportional to $\delta(k_1-k_2)$, so that $\delta(k_1-k_\sigma(N_1)) \delta(k_2-k_\sigma(N_2))$ yields the desired $\delta(N_1-N_2)$.  This is indeed satisfied by the $b$-vacuum $|0_b\rangle$, since every term in the expansion~\eqref{b-vs-a} has zero momentum and the ladder operators in the matrix element cannot survive unless ${\bf k}_1 = -{\bf k}_2$, which of course implies $k_1=k_2$.  More generally, consider any eigenstate $|\Psi\rangle$ of the the total comoving momentum operator $\hat{\bf p}$.  Then using the commutator
\begin{equation}
[\hat{\bf p}, \hat\phi_{{\bf k}_1}(N_1) \hat\phi_{{\bf k}_2}(N_2)] = -({\bf k}_1 + {\bf k}_2) \hat\phi_{{\bf k}_1}(N_1) \hat\phi_{{\bf k}_2}(N_2),
\end{equation}
and taking its expectation value in the state $|\Psi\rangle$, we obtain $({\bf k}_1 + {\bf k}_2) \langle \Psi | \hat\phi_{{\bf k}_1}(N_1) \hat\phi_{{\bf k}_2}(N_2) | \Psi \rangle = 0$.  Therefore, for all eigenstates of the momentum operator (of which $|0_b\rangle$ is a special case), the aforementioned matrix element is proportional to $\delta({\bf k}_1 + {\bf k}_2)$ and hence to $\delta(k_1-k_2)$, thus fulfilling the above condition of whiteness.

The last comment is about the backreaction of the excited modes in non-vaccum states like eq.~\eqref{C2-exp}.  One expects that as we go back to earlier times, these modes blue-shift and eventually backreact on the background spacetime.  Thus in principle we expect that there is a bound on the allowed number of $e$-folds of inflation. Let us begin with a naive analysis of this bound.  A typical two-particle state has comoving energy of order $Q$, so at earlier times the physical energy of the particles grows as $Q/a$, making an increasing contribution to the background energy density and possibly invalidating our description.  Setting $Q/a$ equal to the Planck energy $M_P=1/\sqrt{8\pi G_N}$, we find that the energy of particles exceeds $M_P$ at the scale factor $a_Q(N) = \sigma H/M_P$, i.e., $\log (M_P/\sigma H)$ $e$-folds before $N_Q$ (recall that $N_Q$ is the time that $Q$ crosses the cutoff, as well as the time around which deviation from whiteness becomes maximal).  Thus (for fiducial values $H=10^{-5}M_P$ and $\sigma=10^{-2}$) we can have an interval of about 15 $e$-folds of inflation prior to this type of non-whiteness before running into the troubles of the super-Planckian regime.  However, this naive analysis does not take into account the likelihood of such high-energy pairs and, more importantly, their spatial distribution.  On the contrary, a reasonable answer must involve $\varepsilon$ (so that when $\varepsilon=0$, we fall back to the conventional situation), and must be about the energy density (rather than energy itself --- objects with super-Planckian energy spread across a long distance have no significant backreaction or quantum gravity effects).  We have performed a more careful analysis in appendix~\ref{app:backreaction}, which consists in comparing the energy density of the scalar field with that of the background spacetime.  The final result confirms the qualitative answer of our naive analysis here.


\section{Summary and Conclusions}\label{sec:summary}

We have investigated possible departures from a Gaussian white noise in stochastic inflation for a free test field.  We did so by violating the conventional assumptions that lead to the Gaussian white noise one at a time so that the pure effect of each one is understood.

In section~\ref{sec:white} we noticed that switching from dS to a general FLRW background can only make the amplitude of the white noise time-dependent, without affecting its whiteness or Gaussianity.  So we have a non-stationary Gaussian white noise and the correlator $\langle \xi_\phi(N_1) \xi_\phi(N_2) \rangle$ is of the form of eq.~\eqref{non-stationary-white}, namely, $A(N_1) \delta(N_1-N_2)$.  The amplitude squared $A(N_1)$ of the white noise is related to the dimensionless power spectrum ${\cal P}_\phi$ of field fluctuations according to eq.~\eqref{[phi,phi]+0sharp}.

In section~\ref{sec:window} we studied the effect of the window function by switching from the sharp cutoff, i.e., a step function $W(\kappa)=\theta(\kappa-1)$, to a non-sharp cutoff, which we took for simplicity to be a piecewise liner window function, as depicted in figure~\ref{fig:window-function}.  The width $2\delta$ of this window function controls the time extent $\Delta N_\pm$ of the noise correlation and thus the memory of the stochastic process, as exemplified by eq.~\eqref{DN-vs-delta} and figure~\ref{fig:noise-window} (left).  We found that just adding the width $\delta$ to the window function (but keeping the exact dS background and the vacuum state) doesn't spoil the stationarity of the noise; so in this regard, this is simpler than the case of Subsection~\ref{ssec:white-var}.  However, this time we have a colored noise, with the power spectrum given by eq.~\eqref{power-window-dS} and plotted in figure~\ref{fig:noise-window} (right).  Of course, some of the aspects of our simplistic window function~\eqref{W} are nonphysical, because it lacks the smoothness properties mentioned in ref.~\cite{Winitzki:1999ve}.  Nevertheless the appearance of color is an inevitable feature of a non-sharp cutoff.  More explicitly, we have a stationary Gaussian colored noise, with correlator $\langle \xi_\phi(N_1) \xi_\phi(N_2) \rangle$ of the form of eq.~\eqref{2pt-corr}, i.e., a function (other than the delta function) of the time difference $N_1-N_2$.

In section~\ref{sec:state} we considered a non-Bunch-Davies initial state $|\Psi\rangle$.  One immediate consequence was loss of Gaussianity.  We noticed that in order to have a nontrivial perturbative deviation form white noise, we must have a sum over two-particle states, as in eq.~\eqref{Psi-C2}.  (Incidentally, the widely studied states that differ from the Bunch-Davies by a Bogoliubov transformation cannot produce colored noise, nor can any eigenstate of momentum.  On the other hand, there is a limit on the duration of inflation to avoid backreaction in a non-Bunch-Davies state, which we showed is not severe.)  For a special case that analytical calculation was possible we derived the correlator of the noise in eq.~\eqref{NLO-correlator-Psi}, which is non-stationary and therefore its spectral properties are described by the instantaneous power spectrum --- as opposed to the usual power spectrum --- which we found in eq.~\eqref{inst-power-Psi}.  This provides us with the most general deviation from Gaussian white noise: a non-stationary non-Gaussian colored noise.  It means that the correlator $\langle \xi_\phi(N_1) \xi_\phi(N_2) \rangle$ is not only a function of both $N_1$ and $N_2$, but also higher order correlators are not derivable from the 2-point correlator in the manner of Gaussian variables.

An obvious consequence of a non-white noise is the emergence of memory in the stochastic process.  Therefore, the resulting Lagnevin equation cannot be tackled by the conventional methods and a different Fokker-Planck equation will arise.  There are some methods available in the literature to handle this issue, but it is not the purpose of this paper and we leave such investigations for a future work.

We have restricted our attention to free test fields only. Another source of deviation from Gaussian white noise is the existence of interactions, either with gravity as in the case of inflaton, or with the field itself as in a non-quadratic potential $V(\phi)$. In such cases, one has to include the effect of backreaction, too.  Clearly, this is another source of memory and non-Markovianity that we have not studied in this paper, and can be pursued in the future.


\section*{Acknowledgments}

We would like to thank Farnaz Bahmani, Melika Mirzaei, and Mohammad Hossein Namjoo for useful discussions.  We are thankful to the anonymous referee for very useful remarks.  We also acknowledge financial support from the research council of University of Tehran.  

\appendix

\section{Noise Correlators}\label{app:correlators}

In this appendix we collect a number of expressions for the correlators of the noise operators $\hat\xi_\phi$ and $\hat\xi_v$ given in eqs.~\eqref{xi-phi} and \eqref{xi-v}.  In general, the answers depend on the state $|\Psi\rangle$, the window function $W(\kappa)$, and the background $a(t)$.  We present the results for commutators and anticommutators separately.  In either case, we begin from the most general expressions to the specialized cases.  Below, the spacetime coordinate of an operator are denoted by $x = (t,\b x)$ (or sometimes by $N$ instead of $t$).  The subscripts 1 and 2 indicate whether the expression is evaluated at $t_1$ or $t_2$, and $\Delta \b x = \b x_1 - \b x_2$

\subsection{Commutators}

The commutators of noise operators are independent of the state $|\Psi\rangle$, since the commutator of the creation and annihilation operators is proportional to the identity operator.  Therefore, we have only dependence on $W(\kappa)$ and $a(t)$.  One finds:
\begin{equation}\label{[phi,phi]}
[\hat\xi_\phi(x_1), \hat\xi_\phi(x_2)] = \bar\epsilon_1 \bar\epsilon_2 \int \frac{d^3k}{(2\pi)^3} \left[ \kappa W' \right]_1 \left[ \kappa W' \right]_2 2i\Im \left[ \phi_k(t_1) \phi^*_k(t_2) \right] e^{i\b k \cdot \Delta \b x},
\end{equation}
\begin{equation}\label{[phi,v]}
[\hat\xi_\phi(x_1), \hat\xi_v(x_2)] = \frac{\bar\epsilon_1 \bar\epsilon_2}{H_2} \int \frac{d^3k}{(2\pi)^3} \left[ \kappa W' \right]_1 \left[ \kappa W' \right]_2 2i\Im \left[ \phi_k(t_1) \dot\phi^*_k(t_2) \right] e^{i\b k \cdot \Delta \b x},
\end{equation}
\begin{equation}\label{[v,v]}
[\hat\xi_v(x_1), \hat\xi_v(x_2)] = \frac{\bar\epsilon_1 \bar\epsilon_2}{H_1 H_2} \int \frac{d^3k}{(2\pi)^3} \left[ \kappa W' \right]_1 \left[ \kappa W' \right]_2 2i\Im \left[ \dot\phi_k(t_1) \dot\phi^*_k(t_2) \right] e^{i\b k \cdot \Delta \b x},
\end{equation}
which in the special case of the equal-time commutation relations become:
\begin{align}
&[\hat\xi_\phi(t,\b x), \hat\xi_\phi(t,\b x')] = 0, \qquad
[\hat\xi_v(t,\b x), \hat\xi_v(t,\b x')] = 0, \label{[,]-eqt}\\
&[\hat\xi_\phi(t,\b x), \hat\xi_v(t,\b x')] = \frac{i\bar\epsilon^2}{Ha^3} \int \frac{d^3k}{(2\pi)^3} \left[ \kappa W'(\kappa) \right]^2 e^{i\b k \cdot \Delta \b x}, \label{[phi,v]-eqt}
\end{align}
where use has been made of the Wronskian condition \eqref{Wronskian}.

The preceding results were quite general.  For the sharp cutoff $W(\kappa) = \theta(\kappa-1)$, but otherwise generically (any $a(t)$, hence any mode function $\phi_k$, and any initial state $|\Psi\rangle$), we find, again using the Wronskian:
\begin{equation}\label{[,]-sharp}
[\hat\xi_\phi(x_1), \hat\xi_\phi(x_2)] = 0, \qquad [\hat\xi_v(x_1), \hat\xi_v(x_2)] = 0,
\end{equation}
\begin{equation}\label{[phi,v]-sharp}
[\hat\xi_\phi(x_1), \hat\xi_v(x_2)] = 2i \sigma^3 \bar\epsilon \left( \frac{H}{2\pi} \right)^2 \sinc (k_\sigma |\Delta \b x|) \delta(N_1-N_2),
\end{equation}
where $\sinc(x)=\sin x/x$, and $\epsilon$, $H$ and $k_\sigma$ are evaluated at $N=N_1$.  Incidentally, these equations reveal that the classical behavior emerges in the limit $\sigma\to0$, at least for a sharp cutoff or those approximately equal to it.  For a more careful analysis, see ref.~\cite{Noorbala:2024fim}.

In this paper we use the above relations in the special case of a single patch, which can be trivially obtained by letting $\Delta \b x = 0$.

\subsection{Anticommutators}

Unlike the commutators, the expectation values of the anticommutators do depend on the state $|\Psi\rangle$.  When the criteria of classicality are met, these expectation values can be identified as twice the correlation function of the corresponding classical stochastic variables, i.e., $\langle\Psi| \{ \hat\xi_1, \hat\xi_2 \} |\Psi\rangle \to 2\langle \xi_1 \xi_2 \rangle$.  For generic $|\Psi\rangle$, $W(\kappa)$ and $a(t)$, we have:
\begin{equation}\label{[phi,phi]+}
\langle \{ \hat\xi_\phi(x_1), \hat\xi_\phi(x_2) \} \rangle = \bar\epsilon_1 \bar\epsilon_2 \int \frac{d^3k_1 d^3k_2}{(2\pi)^3} \left[ \kappa W' \right]_1 \left[ \kappa W' \right]_2 2\Re \left[ \langle \hat\phi_{\b k_1}(t_1) \hat\phi_{\b k_2}(t_2) \rangle e^{i\b k_1 \cdot \b x_1 + i\b k_2 \cdot \b x_2} \right],
\end{equation}
\begin{equation}\label{[phi,v]+}
\langle \{ \hat\xi_\phi(x_1), \hat\xi_v(x_2) \} \rangle = \frac{\bar\epsilon_1 \bar\epsilon_2}{H_2} \int \frac{d^3k_1 d^3k_2}{(2\pi)^3} \left[ \kappa W' \right]_1 \left[ \kappa W' \right]_2 2\Re \left[ \langle \hat\phi_{\b k_1}(t_1) \dot{\hat\phi}_{\b k_2}(t_2) \rangle e^{i\b k_1 \cdot \b x_1 + i\b k_2 \cdot \b x_2} \right],
\end{equation}
\begin{equation}\label{[v,v]+}
\langle \{ \hat\xi_v(x_1), \hat\xi_v(x_2) \} \rangle = \frac{\bar\epsilon_1 \bar\epsilon_2}{H_1H_2} \int \frac{d^3k_1 d^3k_2}{(2\pi)^3} \left[ \kappa W' \right]_1 \left[ \kappa W' \right]_2 2\Re \left[ \langle \dot{\hat\phi}_{\b k_1}(t_1) \dot{\hat\phi}_{\b k_2}(t_2) \rangle e^{i\b k_1 \cdot \b x_1 + i\b k_2 \cdot \b x_2} \right].
\end{equation}

Further simplification can be achieved if we pick the vacuum state $|\Psi\rangle = |0\rangle$ (annihilated by all $\hat a_{\bf k}$s).  Then for generic window function $W(\kappa)$ and background $a(t)$ (hence any mode function $\phi_k$), we have:
\begin{equation}\label{[phi,phi]+0}
\langle \{ \hat\xi_\phi(x_1), \hat\xi_\phi(x_2) \} \rangle_0 = \bar\epsilon_1 \bar\epsilon_2 \int \frac{d^3k}{(2\pi)^3} \left[ \kappa W' \right]_1 \left[ \kappa W' \right]_2 2\Re \left[ \phi_k(t_1) \phi^*_k(t_2) \right] e^{i\b k \cdot \Delta \b x},
\end{equation}
\begin{equation}\label{[phi,v]+0}
\langle \{ \hat\xi_\phi(x_1), \hat\xi_v(x_2) \} \rangle_0 = \frac{\bar\epsilon_1 \bar\epsilon_2}{H_2} \int \frac{d^3k}{(2\pi)^3} \left[ \kappa W' \right]_1 \left[ \kappa W' \right]_2 2\Re \left[ \phi_k(t_1) \dot\phi^*_k(t_2) \right] e^{i\b k \cdot \Delta \b x},
\end{equation}
\begin{equation}\label{[v,v]+0}
\langle \{ \hat\xi_v(x_1), \hat\xi_v(x_2) \} \rangle_0 = \frac{\bar\epsilon_1 \bar\epsilon_2}{H_1 H_2} \int \frac{d^3k}{(2\pi)^3} \left[ \kappa W' \right]_1 \left[ \kappa W' \right]_2 2\Re \left[ \dot\phi_k(t_1) \dot\phi^*_k(t_2) \right] e^{i\b k \cdot \Delta \b x}.
\end{equation}
If we in addition assume a sharp cutoff $W(\kappa) = \theta(\kappa-1)$, we find
\begin{equation}\label{[phi,phi]+0sharp}
\langle \{ \hat\xi_\phi(x_1), \hat\xi_\phi(x_2) \} \rangle_0 = 2\bar\epsilon {\cal P}_\phi \sinc(k_\sigma |\Delta \b x|) \delta(N_1-N_2),
\end{equation}
\begin{equation}\label{[phi,v]+0sharp}
\langle \{ \hat\xi_\phi(x_1), \hat\xi_v(x_2) \} \rangle_0 = 2\bar\epsilon {\cal P}_{\phi,v} \sinc(k_\sigma |\Delta \b x|) \delta(N_1-N_2),
\end{equation}
\begin{equation}\label{[v,v]+0sharp}
\langle \{ \hat\xi_v(x_1), \hat\xi_v(x_2) \} \rangle_0 = 2\bar\epsilon {\cal P}_v \sinc(k_\sigma |\Delta \b x|) \delta(N_1-N_2),
\end{equation}
where 
\begin{equation}
{\cal P}_f(k,N) = \frac{k^3}{2\pi^2} |f_k(N)|^2
\end{equation}
is the dimensionless power spectrum\footnote{Notice the distinction between this power spectrum of the field $\phi(\b x)$ living in $\mathbb{R}^3$ and the power spectrum of the noise $\xi_\phi(N)$ that lives in $\mathbb{R}$.  See the footnote~\ref{ft:power} in appendix~\ref{app:noise} on the distinction between $d=3$ and $d=1$.\label{ft:PvsP}} for $f=\phi$ or $f=v=d\phi/dN$, and ${\cal P}_{\phi,v} = \frac{k^3}{2\pi^2} \Re(\phi v^*)$, all of which are evaluated at $k=k_\sigma(N_1)$ and $N=N_1$.  

A final simplification occurs (still in $|0\rangle$ and with the sharp cutoff) in exact dS background, where the mode functions and power spectra are known.  For the massless case:
\begin{equation}\label{[phi,phi]+0sharpdSm=0}
\langle \{ \hat\xi_\phi(x_1), \hat\xi_\phi(x_2) \} \rangle_0 = 2 (1+\sigma^2) \left( \frac{H}{2\pi} \right)^2 \sinc(k_\sigma |\Delta \b x|) \delta(N_1-N_2),
\end{equation}
\begin{equation}\label{[phi,v]+0sharpdSm=0}
\langle \{ \hat\xi_\phi(x_1), \hat\xi_v(x_2) \} \rangle_0 = -2 \sigma^2 \left( \frac{H}{2\pi} \right)^2 \sinc(k_\sigma |\Delta \b x|) \delta(N_1-N_2),
\end{equation}
\begin{equation}\label{[v,v]+0sharpdSm=0}
\langle \{ \hat\xi_v(x_1), \hat\xi_v(x_2) \} \rangle_0 = 2 \sigma^4 \left( \frac{H}{2\pi} \right)^2 \sinc(k_\sigma |\Delta \b x|) \delta(N_1-N_2).
\end{equation}
For the massive case, under the assumption
\begin{equation}\label{sigma-interval}
\exp \left(-\frac{3H^2}{m^2} \right) \ll \sigma^2 \ll \frac{m^2}{3H^2} \ll 1,
\end{equation}
the leading terms become
\begin{equation}\label{[phi,phi]+0sharpdS}
\langle \{ \hat\xi_\phi(x_1), \hat\xi_\phi(x_2) \} \rangle_0 = 2\left( \frac{H}{2\pi} \right)^2 \sinc(k_\sigma |\Delta \b x|) \delta(N_1-N_2),
\end{equation}
\begin{equation}\label{[phi,v]+0sharpdS}
\langle \{ \hat\xi_\phi(x_1), \hat\xi_v(x_2) \} \rangle_0 = -2\frac{m^2}{3H^2} \left( \frac{H}{2\pi} \right)^2 \sinc(k_\sigma |\Delta \b x|) \delta(N_1-N_2),
\end{equation}
\begin{equation}\label{[v,v]+0sharpdS}
\langle \{ \hat\xi_v(x_1), \hat\xi_v(x_2) \} \rangle_0 = 2\left( \frac{m^2}{3H^2} \right)^2 \left( \frac{H}{2\pi} \right)^2 \sinc(k_\sigma |\Delta \b x|) \delta(N_1-N_2).
\end{equation}

As with the commutators, it's again a simple matter to obtain the results for the anticommutators in a single patch by letting $\Delta \b x = 0$.

\subsection{Higher Order Correlators}

In general, higher order correlation functions are needed to completely specify the stochastic process.  For a generic state $|\Psi\rangle$, this is complicated.  But for the vacuum state $|0\rangle$, it is easy to show that the resulting classical process is Gaussian, as we do here.

Since the noise operators generally do not commute at different points, $x_i$, the $n$-point correlator depends on the ordering of fields.  If a classical picture is to emerge, however, this ordering must be irrelevant.  Therefore, let us consider the time-ordered product which satisfies Wick's theorem:
\begin{equation}\label{Wick}
	\langle {\rm T} \{ \hat\xi_\phi(x_1) \hat\xi_\phi(x_2) \ldots \hat\xi_\phi(x_n) \} \rangle_0 = \sum_\text{pairings} \prod \langle {\rm T} \{ \hat\xi_\phi(x_i) \hat\xi_\phi(x_j) \} \rangle_0.
\end{equation}
Barring the time-ordering symbol, this is precisely the requirement for a classical stochastic process to be Gaussian.  Note that the two-point correlators on the right hand side need not contain a delta function $\delta(N_i-N_j)$; so this result is equally valid for colored and white noise.


\section{Review of the Mathematical Properties of Noise}\label{app:noise}

In this appendix we present a self-contained review of required definitions of noise properties for a generic stochastic process $f(t)$.\footnote{For applications in stochastic inflation, the relevant random processes are the noise $\xi_\phi(t)$ and the coarse-grained field $\phi_l(t)$ which, apart from the drift term, is the integrated noise.  The results of this appendix can be readily generalized to more general random fields where $t$ is replaced by a $d$-dimensional variable $\b x\in\mathbb{R}^d$.  They can then be used for the other important application in cosmology, namely, the cosmological perturbations, where the random field is, e.g., the initial curvature perturbation ${\cal R}({\bf x})$ with ${\bf x}\in\mathbb{R}^3$.  Also we assume that $f$ is real as is in our use case, but extension to complex $f$ whose real and imaginary parts are independent processes is straightforward.}  For more details on some of these, see ref.~\cite{vanKampen}.  The process $f(t)$ is completely determined, once the joint probability $P[f(t_1),\ldots,f(t_n)]$ of values of $f$ is specified for all possible choices of the $n$ points $t_1,\ldots,t_n$. Equivalently, one may specify all $n$-point correlators $\langle f(t_1) \ldots f(t_n) \rangle$.  

If $f$ is \textit{independent} at different times, i.e., if $P[f(t_1),\ldots,f(t_n)] = P[f(t_1)] \ldots P[f(t_n)]$, then all correlators factorize:
\begin{equation}\label{independence}
\langle f(t_1) \ldots f(t_n) \rangle = \langle f(t_1) \rangle \ldots \langle f(t_n) \rangle.
\end{equation}
Consequently, the integrated process $F(t) = \int f(t)dt$ will be a Markov process and has no memory.  In other words, by interpreting such an $f$ as a \textit{noise} whose accumulations make up the process $F$, we find that the noise $f$ (being independent at different times) induces no memory on the process $F$.
Equivalently, one says that the increments of a Markov process are independent.

A distinct notion of interest to us is stationarity: $f$ is called \textit{stationary}\footnote{In other contexts where $t$ is replaced by a $d$-dimensional spatial variable $\b x \in \mathbb{R}^d$, we call it a \textit{homogeneous} random field rather than a stationary process.  In the same context, invariance of the correlators under rotations of the coordinates yields an \textit{isotropic} random field.} if $\langle f(t_1) \ldots f(t_n) \rangle$ is a function only of the time differences, so that
\begin{equation}
\langle f(t_1+t) \ldots f(t_n+t) \rangle = \langle f(t_1) \ldots f(t_n) \rangle.
\end{equation}
It then follows that the correlators of the Fourier transform,
\begin{equation}
{\widetilde f}(\omega) = \frac{1}{\sqrt{2\pi}} \int_{-\infty}^\infty f(t) e^{i\omega t} dt,
\end{equation}
satisfy
\begin{equation}
\langle \widetilde f(\omega_1) \ldots \widetilde f(\omega_n) \rangle \propto \delta(\omega_1 + \ldots + \omega_n).
\end{equation}
For a stationary process, the \textit{power spectrum} $P_f$ is defined as the coefficient of proportionality when $n=2$,
\begin{equation}\label{P_f}
\langle \widetilde f(\omega) \widetilde f(\omega') \rangle = P_f(\omega) \delta(\omega + \omega'),
\end{equation}
which is related to the two-point correlator
\begin{equation}\label{2pt-corr}
C_f(t) = \langle f(0) f(t) \rangle = \langle f(t) f(0) \rangle = C_f(-t)
\end{equation}
via Fourier transformation: $P_f(\omega) = \sqrt{2\pi} \widetilde{C_f}(\omega)$.  The evenness of $C_f$ in $t$ implies that $P_f$ is an even function of $\omega$; and the reality of $f$ implies, via eq.~\eqref{P_f}, that $P_f$ is a non-negative real function of $\omega$.  The shape of the power spectrum characterizes the spectral properties, or for short, the \textit{color} of $f$.  In particular, for a white noise, $P_f(\omega)$ is a constant independent of $\omega$, with standard normalization corresponding to $P_f(\omega)=1$.  Other profiles have their own names too, e.g., pink noise for $P_f(\omega)\propto1/\omega$, etc.

The physical meaning of the power is made clear if $f$ is regarded as an electric current source.  Then from Parseval's theorem, the expected total energy (assuming unit resistance) of the source is $\langle \int_{-\infty}^\infty f(t)^2 dt \rangle = \langle \int_{-\infty}^\infty \widetilde f(\omega) \widetilde f(-\omega) d\omega \rangle$, which justifies the name \textit{energy spectrum} (expected energy per angular frequency interval) for $\langle |\widetilde f|^2 \rangle$.  However, $\langle \int_{-\infty}^\infty f^2 dt \rangle$ diverges for a stationary process since $\langle f^2 \rangle$ is $t$-independent, or alternatively, because of the $\delta(0)$ in eq.~\eqref{P_f}.  Therefore, the appropriate quantity is not the expected total energy, rather the expected total power: $\lim_{T\to\infty} \frac1{T} \langle \int_{-T/2}^{T/2} f(t)^2 dt \rangle$.  After regularization, the $\delta(0)$ in eq.~\eqref{P_f} becomes $1/\Delta\omega$ (where $\Delta\omega = 2\pi/T$ is the smallest allowed frequency for finite $T$) and cancels the $T$ in the denominator.  Thus we find $\int_{-\infty}^\infty P_f \frac{d\omega}{2\pi}$ as the expected total power, and $\frac1{2\pi} P_f$ as the expected power per angular frequency interval.  This is equivalent to saying that $P_f$ is the expected power per frequency interval, or simply the power spectrum.

The notion of power spectrum can be extended to non-stationary processes in several ways.  We begin by investigating the physical meaning of power in the electric current analogy, \`a la ref.~\cite{Page}.  Consider, instead of the expected total energy, the expected energy supplied by the source from the infinite past up to time $t_0$, i.e., $E_f^{-}(t_0) = \langle \int_{-\infty}^{t_0} f(t)^2 dt \rangle = \langle \int_{-\infty}^\infty |\widetilde{f^-}(\omega)|^2 d\omega \rangle$, where $f^{-}(t) = f(t) \theta(t_0-t)$.  At time $t_0$, the instantaneous power supplied by the source is $dE_f^{-}(t_0)/dt_0$, which can be easily shown to be equal to $\int_{-\infty}^\infty P_f^{-}(\omega,t_0) \frac{d\omega}{2\pi}$, where
\begin{equation}
P_f^{-}(\omega,t_0) = 2\sqrt{2\pi} \Re \left[ \langle f(t_0) \widetilde{f^-}(\omega) \rangle e^{-i\omega t_0} \right] = 2 \Re \int_{-\infty}^0 \langle f(t_0) f(t_0+t)\rangle e^{i\omega t} dt
\end{equation}
is the instantaneous expected power per frequency interval.  Clearly, $\int_{-\infty}^\infty P_f^{-}(\omega,t_0) \frac{d\omega}{2\pi}$ is equal to $\langle f(t_0)^2 \rangle$, which is consistent with what we expect from $dE_f^{-}(t_0)/dt_0$, and is positive.  However, the nonintegrated $P_f^-(\omega,t_0)$ has only the evenness in $\omega$ and the reality properties, but is not necessarily positive, which indicates that the contribution of a certain frequency interval to the power may decrease in some periods of time.  We can similarly construct $E_f^{+}(t_0) = \langle \int^{\infty}_{t_0} f(t)^2 dt \rangle$ and $f^{+}(t) = f(t) \theta(t-t_0)$ to compute $-dE_f^{+}(t_0)/dt_0$, which yields:
\begin{equation}
P_f^{+}(\omega,t_0) = 2\sqrt{2\pi} \Re \left[ \langle f(t_0) \widetilde{f^+}(\omega) \rangle e^{-i\omega t_0} \right] = 2 \Re \int^{\infty}_0 \langle f(t_0) f(t_0+t)\rangle e^{i\omega t} dt.
\end{equation}
Evidently, the sum of $E_f^+$ and $E_f^-$ is $t_0$-independent, so the total power evaluated by either of them is the same.  However, the power spectra $P_f^+$ and $P_f^-$ are different, because of the effect introduced by the cutoff $t_0$ on the frequency content of $dE_f^\pm/dt_0$.  In addition, an average instantaneous power spectrum is defined by:
\begin{equation}
\frac12 \left( P_f^+ + P_f^- \right) = \Re \bar P_f (\omega,t_0),
\end{equation}
where
\begin{equation}
\bar P_f (\omega,t_0) = \sqrt{2\pi} \langle f(t_0) \widetilde{f}(\omega) \rangle e^{-i\omega t_0} = \int_{-\infty}^\infty \langle \widetilde f(\omega') \widetilde f(\omega) \rangle e^{-i(\omega+\omega')t_0} d\omega'.
\end{equation}
Obviously, $\bar P_f$ is related to the Fourier transform of the correlator
\begin{equation}
\bar C_f(t,t_0) = \langle f(t_0) f(t_0+t) \rangle = \langle f(t_0+t) f(t_0) \rangle = \bar C_f(-t,t_0+t)
\end{equation}
with respect to $t$, namely,
\begin{equation}
\bar P_f(\omega,t_0) = \sqrt{2\pi} \widetilde{\bar C_f}(\omega,t_0) = \int_{-\infty}^\infty \langle f(t_0) f(t_0+t)\rangle e^{i\omega t} dt.
\end{equation}

The $\bar P_f$ defined above has the disadvantage that it is in general a complex number.  Now consider a symmetric form of the correlator at a fixed time $t_0$, i.e.,
\begin{equation}
C_f(t,t_0) = \langle f(t_0-\frac{t}2) f(t_0+\frac{t}2) \rangle = \langle f(t_0+\frac{t}2) f(t_0-\frac{t}2) \rangle = C_f(-t,t_0),
\end{equation}
which is even in $t$.  We can then define the \textit{instantaneous power spectrum} of a non-stationary process $f$ at time $t_0$ by Fourier transforming $C_f(t,t_0)$ with respect to $t$:
\begin{equation}\label{inst-power}
\begin{aligned}
P_f(\omega,t_0) &= \sqrt{2\pi} \widetilde{C_f}(\omega,t_0) \\
&= \int_{-\infty}^\infty \langle f(t_0-\frac{t}2) f(t_0+\frac{t}2) \rangle e^{i\omega t} dt = \int_{-\infty}^\infty \langle \widetilde f(\frac{\omega'}2-\omega) \widetilde f(\frac{\omega'}2+\omega) \rangle e^{-i\omega't_0} d\omega'.
\end{aligned}
\end{equation}
As before, the evenness of $C_f$ in $t$ implies that $P_f$ is an even function of $\omega$.  The reality of $f$ implies that $P_f(\omega,t_0) = P_f(-\omega,t_0)^*$, from which we deduce that $P_f(\omega,t_0)$ is real.  So $P_f$ is nicer than $\bar P_f$.  But note that unlike the stationary case, $P_f(\omega,t_0)$ is not necessarily positive.  In fact, $P_f$ is, apart from the expectation value, the Wigner transformation of $f$, which is well known to have this negativity property.\footnote{The Wigner distribution of the pure state $|\psi\rangle$ is defined as $W(x,p) = \int_{-\infty}^\infty \langle x-\frac{y}{2} | \psi \rangle \langle \psi | x+\frac{y}{2} \rangle e^{iyp} dy$.}  The analogy with quantum mechanics goes even further: $W(x,p)$, being the (possibly negative) classical phase space probability distribution that corresponds to the quantum state $|\psi\rangle$, simultaneously assigns a position and a momentum to the particle.  In the same manner, $P_f(\omega,t_0)$ simultaneously assigns a time and a frequency to the signal $f$.  In either case, there is no way to simultaneously determine the noncommuting pairs (position/momentum or time/frequency) in a perfect manner, thus ending up with negative values of the distribution.  We also have the marginal property
\begin{equation}
\int_{-\infty}^\infty P_f(\omega,t_0) \frac{d\omega}{2\pi} = \langle f(t_0)^2 \rangle, \qquad \int_{-\infty}^\infty P_f(\omega,t_0) dt_0 = \langle |\widetilde f(\omega)|^2 \rangle,
\end{equation}
and similarly for the quantum mechanical problem.

It is clear that $C_f$, $\bar C_f$, $P_f$ and $\bar P_f$ all convey the same information and any one of them can be found from any other by Fourier transformation and/or $\bar C_f(t,t_0) = C(t,t_0+\frac{t}2)$.  $P^\pm_f$ can be obtained from them, too.  Any of the functions $P_f$, $\bar P_f$, or $P^\pm_f$ may be called the instantaneous power spectrum, but here we reserve this term for $P_f$ and exclusively work with it.\footnote{In the signal processing literature, $P^-_f$ is attributed to Page~\cite{Page} and $P^+_f$ to Levin~\cite{Levin}, $\bar P_f$ is called the Rihaczek distribution~\cite{Rihaczek}, and $P_f$ is known as the Wigner-Ville distribution.  For a review, see ref.~\cite{de-Oliveira}.}  All of them are even functions of $\omega$; and except for $\bar P_f$, they are all real.  In the stationary case, $\bar P_f$ becomes real too, and all of them become $t_0$-independent and coincide with our previous definition of power in eq.~\eqref{P_f}.  

In a manner similar to stationarity, one defines scale-invariance of a process.  Suppose $f$ has time dimension $\Delta$ (for example, if $f$ is a frequency, $\Delta=-1$).  Then we say that $f$ is \textit{scale-invariant}, if
\begin{equation}
\langle f(\lambda t_1) \ldots f(\lambda t_n) \rangle = \lambda^{n\Delta} \langle f(t_1) \ldots f(t_n) \rangle.
\end{equation}
In terms of $\widetilde f$:
\begin{equation}
\langle \widetilde f(\lambda^{-1} \omega_1) \ldots \widetilde f(\lambda^{-1} \omega_n) \rangle = \lambda^{n(\Delta+d)} \langle f(\omega_1) \ldots f(\omega_n) \rangle,
\end{equation}
which is consistent with the fact that, in the more general case $t\in\mathbb{R}^d$, $\widetilde f$ has time dimension $\Delta+d$.  It follows that if a process is both stationary and scale-invariant then its power spectrum will be of the form
\begin{equation}\label{scale-inv-pow}
P_f(\omega) = P_f(\omega_0) \left( \frac{\omega_0}{\omega} \right)^{2\Delta+d}.
\end{equation}
The white noise falls in this class, with $\Delta=-1/2$ and $d=1$.\footnote{The well-known scale-invariant cosmological curvature perturbation ${\cal R}({\bf x})$ also falls in this class, with $\Delta=0$ (since $\cal R$ is dimensionless), $d=3$ (since ${\bf x}\in\mathbb{R}^3)$ and $P_{\cal R}({\bf k})\propto k^{-3}$ (in the common notation where $k$ is used instead of $\omega$).  The conventional definition of the dimensionless power spectrum ${\cal P}_{\cal R} = k^3 P_{\cal R}/2\pi^2$ is also a consequence of eq.~\eqref{scale-inv-pow} which implies that $\omega^{2\Delta+d}P_f$ is $\omega$-independent.\label{ft:power}}

While reviewing the definitions above, we alluded to the main property of the standard white noise, namely,
\begin{equation}\label{<WW>-time}
\langle \xi_n(t) \xi_n(t') \rangle = \delta(t-t'),
\end{equation}
or equivalently,
\begin{equation}\label{<WW>-freq}
\langle \widetilde{\xi_n}(\omega) \widetilde{\xi_n}(\omega') \rangle = \delta(\omega + \omega').
\end{equation}
The subscript $n$ is to emphasize the normalization, and we call it a \textit{standard} white noise.  Any multiple of $\xi_n$, like $f=\alpha\xi_n$ is also called a (non-standard) white noise with power $P_f=\alpha^2$, if $\alpha$ is a constant independent of $t$.  If the multiplicative factor $\alpha(t)$ is $t$-dependent, we no longer have a stationary process, but since the instantaneous power $P_f(\omega,t_0)=\alpha(t_0)^2$ is still flat (independent of $\omega$), one still calls it a (non-stationary) white noise with time-dependent normalization.  More generally, if $f$ is uncorrelated at separate times, i.e.,
\begin{equation}\label{non-stationary-white}
\langle f(t) f(t') \rangle = A(t) \delta(t-t'),
\end{equation}
then $P_f(\omega,t_0) = A(t_0)$ becomes $\omega$-independent, which is again a white noise with a time-dependent normalization.  In such cases, it is possible to use a new time variable $\bar t = \int dt/\sqrt{A(t)}$ (so that $A(t) \delta(t-t') = \delta(\bar t-\bar {t'})$), in terms of which $f$ appears as a standard white noise.

Eqs.~\eqref{<WW>-time} and \eqref{<WW>-freq} in the preceding paragraph cannot constitute a complete specification of a stochastic process, as the information about higher correlators beyond two-point are missing.  To complete the definition of a \textit{Gaussian} standard white noise, we amend eq.~\eqref{<WW>-time} by requiring that all higher order correlators are completely determined by the two-point correlator via the Wick theorem, e.g.,
\begin{equation}
\langle \xi_n(t_1) \xi_n(t_2) \xi_n(t_3) \xi_n(t_4) \rangle = \langle \xi_n(t_1) \xi_n(t_2) \rangle \langle \xi_n(t_3) \xi_n(t_4) \rangle + \ldots = \delta(t_1-t_2) \delta(t_3-t_4) + \ldots
\end{equation}
and so on (of course, all odd correlators vanish, including $\langle \xi_n \rangle$).  We can equivalently, and more concisely, write the definition of Gaussian white noise as
\begin{equation}\label{def:gaussian-white}
\langle \xi(t_1) \ldots \xi(t_k) \rangle_c = 0, \qquad \text{for all $k>2$},
\end{equation}
where the subscript $c$ denotes cumulant (i.e., connected diagrams).  A \textit{non-Gaussian} white noise fails eq.~\eqref{def:gaussian-white} but satisfies
\begin{equation}
\langle \xi(t_1) \ldots \xi(t_k) \rangle_c = A_k \prod_{i=1}^{k-1} \delta(t_i-t_{i+1}), \qquad \text{for all $k\geq1$}.
\end{equation}
In a stationary process, $A_k$s are constants; but more generally they could depend on time.  Note that in either case, for distinct $t_i$s, even a non-Gaussian white noise satisfies eq.~\eqref{independence} and hence induces no memory.  In words, whiteness requires independence at different times and memorylessness; Gaussianity requires the vanishing of the cumulants beyond the second.

%
%
%


\section{Green's Method for Mode Functions with Slowly Varying Hubble}\label{app:Green}

Here we develop an approximation to calculate the solution of the Mukhanov-Sasaki equation for the mode function in an inflationary background that is not exactly dS.  We denote by $\bar{u}(\tau)$ the unperturbed solution in the exact dS case.  Assuming the deviation from dS is small, we express the solution as\footnote{We have dropped the subscript $k$ on $u_k$ in this appendix to avoid clutter in notation.}
\begin{equation}
u(\tau) = \bar{u}(\tau) + \delta u(\tau),
\end{equation}
where
\begin{equation}\label{ubar}
\bar{u} = \frac{-1}{2}\sqrt{-\pi \tau}H_{\nu_{0}}(-k\tau)
\end{equation}
with the time-independent parameter $\nu_0 = \frac32$.\footnote{We could also set $\nu_0 = \sqrt{\frac{9}{4}-\frac{m^2}{H_{0}^2}}$ with $H_0$ the value of the Hubble at some time during inflation.  That would add an irrelevant phase to eq.~\eqref{ubar} and complicate some of the subsequent equations without much gain, unless $m/H$ is appreciable.} Thus $\delta u$ quantifies the deviation induced by the dynamics of the Hubble parameter $H$. The full mode function $u(\tau)$ satisfies the Mukhanov-Sasaki equation, which we can write as
\begin{equation}
u'' + \left( k^2-\frac{\nu(\tau)^2-\frac14}{\tau^2} \right) u = 0,
\label{MS-nu}
\end{equation}
where
\begin{equation}
\nu^2(\tau)=\nu_{0}^2+\delta\nu^2(\tau).
\label{delta-nu}
\end{equation}
By plugging $u$ into eq.~\eqref{MS-nu} and using the background equation for $\bar u$, as well as eq.~\eqref{delta-nu}, we have, to leading order in perturbations:
\begin{equation}
\label{delta-u-eq}
\delta u'' + \left( k^2-\frac{\nu_{0}^{2}-1/4}{\tau^2} \right) \delta u=\frac{\delta\nu(\tau)^2}{\tau^2}\bar{u}.
\end{equation}
The boundary conditions to be used are $\delta u = \delta u' = 0$ at $\tau=-\infty$, since $\bar u$ already satisfies the asymptotic condition $u \to e^{-ik\tau}/\sqrt{2k}$.

Eq.~\eqref{delta-u-eq} can be solved using Green's method as follows:
\begin{equation}
\label{Green-method}
\delta u(\tau)=\int_{-\infty}^{\infty} G(\tau,\tau') S(\tau')d\tau',
\end{equation}
where the source term $S(\tau) = \delta\nu^2\bar{u}/\tau^2$ is the right hand side of eq.~\eqref{delta-u-eq}, and $G$ is the Green's function of the differential operator on the left hand side.  The most general form for the Green's function in eq.~\eqref{Green-method} is
\begin{equation}
G(\tau,\tau')=\sqrt{-k\tau}
\begin{cases}
A_{1}^{+}(\tau')H_{\nu_{0}}(-k\tau)+A_{2}^{+}(\tau')H_{\nu_{0}}^{*}(-k\tau) & \tau>\tau', \\
A_{1}^{-}(\tau')H_{\nu_{0}}(-k\tau)+A_{2}^{-}(\tau')H_{\nu_{0}}^{*}(-k\tau) & \tau<\tau'. \\
\end{cases}
\end{equation}
In the limit $\tau\rightarrow -\infty$, we impose the boundary conditions $\delta u=\delta u'=0$, which imply:
\begin{equation}
G(\tau\rightarrow-\infty,\tau')= \frac{\partial G}{\partial\tau}(\tau\rightarrow-\infty,\tau')=0.
\end{equation}
Consequently, we find $A_{1}^{-}=A_{2}^{-}=0$, which means that $G(\tau,\tau')$ is proportional to $\theta(\tau-\tau')$, so the upper limit of integration in eq.~\eqref{Green-method} can be replaced by $\tau$.  Now by taking into account the continuity condition of $G$ and discontinuity of its derivative at $\tau=\tau'$, namely,
\begin{equation}
\left. G(\tau,\tau') \right|_{ \tau'+} = \left. G(\tau,\tau') \right|_{ \tau'-}, \qquad
\left. \frac{\partial G(\tau,\tau')}{\partial \tau} \right|_{ \tau'+} - \left. \frac{\partial G(\tau,\tau')}{\partial \tau} \right|_{ \tau'-} = 1,
\end{equation}
together with the Wronskian relation for Hankel functions,\footnote{$H_{\nu}(x) H'^*_{\nu}(x) - H'_{\nu}(x) H_{\nu}^*(x) = -4i/\pi x$ with prime meaning $d/dx$ here.} we can fix the other two coefficients:
\begin{equation}
A_{1}^{+}(\tau') = A_2^{+}(\tau')^* = \frac{i \pi}{4k}\sqrt{-k\tau'} H_{\nu_{0}}^{*}(-k\tau').
\end{equation}
Thus the Green's function $G(\tau,\tau')$ can finally be expressed as
\begin{equation}\label{Green}
\begin{aligned}
G(\tau ,\tau') &= \frac{i \pi\sqrt{\tau\tau'}}{4} \left[ H_{\nu_{0}}(-k\tau) H_{\nu_{0}}^{*}(-k\tau') - H_{\nu_{0}}^{*}(-k\tau) H_{\nu_{0}}(-k\tau') \right] \theta(\tau-\tau') \\
&= \frac{1}{k^3 \tau \tau'} \left[ (k^2 \tau \tau' + 1 ) \sin k (\tau-\tau') - k (\tau-\tau') \cos k (\tau-\tau') \right] \theta(\tau-\tau'),
\end{aligned}
\end{equation}
where $\nu_0=\frac32$ is explicitly used in the second line.  Therefore, eqs.~\eqref{Green-method} and \eqref{Green} provide the solution to eq.~\eqref{delta-u-eq} as
\begin{equation}\label{du-Green}
\delta u(\tau) = \int_{-\infty}^{\tau} \frac{(k^2 \tau \tau' + 1 ) \sin k (\tau-\tau') - k (\tau-\tau') \cos k (\tau-\tau')}{k^3 \tau \tau'} \frac{\delta\nu(\tau')^2}{\tau'^2} \frac{e^{-ik\tau'}}{\sqrt{2k}} \left( 1 - \frac{i}{k\tau'} \right) d\tau'.
\end{equation}

In what follows, we are going to calculate $\delta \nu^{2}$ for a spectator field $\phi$. The Mukhanov-Sasaki equation for such a field is given by eq.~\eqref{MS}.  Comparing this with eq.~\eqref{MS-nu} we conclude
\begin{equation}\label{nu2-test}
\nu^{2} = \left( \frac{a''}{a} - m^2a^2 \right) \tau^{2} + \frac{1}{4};
\end{equation}
and finally from \eqref{delta-nu} we obtain for $\nu_0=\frac32$:
\begin{equation}\label{dnu2-3/2}
\delta \nu^{2} = \left( \frac{a''}{a} - m^2a^2 \right) \tau^{2} - 2 = 2 [ (aH\tau)^2 - 1] - \epsilon (aH\tau)^2 - (ma\tau)^2.
\end{equation}
If the evolution of the scale factor is given, one can in principle calculate $\delta\nu^2$ in terms of $\tau$ and plug in eq.~\eqref{du-Green} to obtain $\delta u$.  We will do this below for a toy model.  But for the sake of completeness, let us write an approximation for $\delta\nu^2$ that can be used in the generic case.  First, recall that in an exact dS background we have $aH\tau=-1$ (upon choosing $\tau=0$ at future infinity and a suitable normalization for $a$), which yields $\delta\nu^2 = -(m/H)^2$.  But in general, we have
\begin{equation}
\frac{d}{d\tau}\left( \frac{1}{aH}\right) = \epsilon -1 \Longrightarrow \frac{1}{aH} = -\tau - \int^{\tau_e}_\tau \epsilon(\tau')d\tau',
\end{equation}
where $\tau_e$ is an arbitrary moment whose conformal time we choose to satisfy $\tau_e = -1/(aH)|_e$.  To make the integral term small compared to $-\tau$, we take $\tau_e$ to be about the end of slow-roll era, e.g., when $\epsilon=0.1$.  Therefore, to the leading order in $\epsilon$ and $m/H$, we have
\begin{equation}\label{delta-nu2}
\delta\nu^2 = - \left( \frac{m}{H} \right)^2 - \epsilon + \frac4{-\tau} \int^{\tau_e}_\tau \epsilon(\tau')d\tau'.
\end{equation}
This form would be convenient if only the leading slow-roll contributions are required.

\begin{figure}
     \centering
     \includegraphics[width=0.6\textwidth]{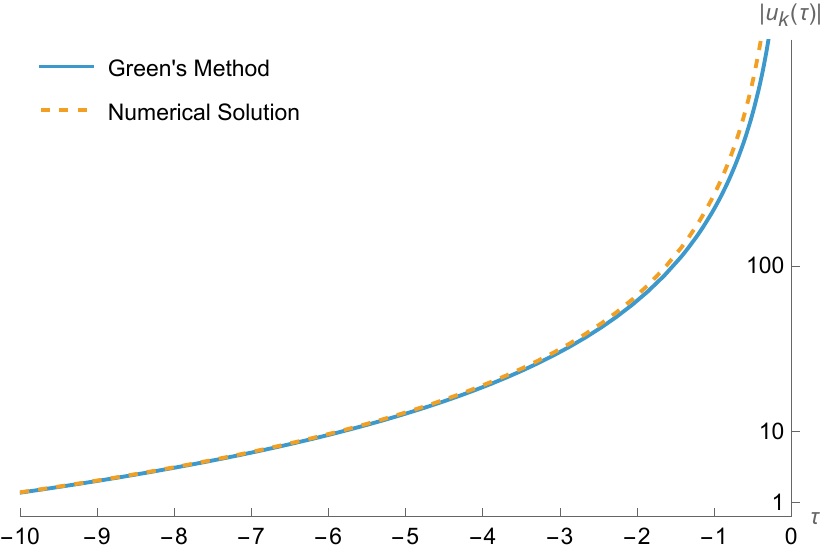}
     \caption{The plot of $|u_k|$ for the toy model ~\eqref{toy-model} with $c=1$ and $k=0.01 H_0$, using Green's method (solid blue) and the numerical solution (dashed orange). The vertical and horizontal axes have units of $1/\sqrt{2k}$ and $1/H_0$, respectively. }
     \label{fig:green-method}
 \end{figure}

To show how things work, we consider a toy model in which $m=0$ and the scale factor behaves as
\begin{equation}
a(\tau) = -\frac{1}{H_0\tau} + \frac{c}{H_0^2\tau^2},
	\label{toy-model}
\end{equation}
where $\tau$ is the conformal time and $c$ and $H_0$ are positive constants.  As $\tau$ varies from $-\infty$ to $0$, the dominant term switches from the first term to the second one, featuring a transition from dS to a universe filled with a perfect fluid with the equation of state parameter $w=-2/3$.   The Hubble parameter
\begin{equation}
H = \frac{H_0\tau - 2c}{(H_0\tau - c)^2} H_0^2\tau \approx H_0 \left( 1 - \frac{c^2}{H_0^2\tau^2} \right)
\end{equation}
is clearly time-dependent.  (Here and below, the last approximation is valid for early times.)  This is not a realistic inflationary model with graceful exist; in fact, although the slow-roll parameter
\begin{equation}
\epsilon = \frac{2c^2}{(H_0\tau-2c)^2} \approx \frac{2c^2}{H_0^2\tau^2} \left( 1 + \frac{4c}{H_0\tau} \right)
\end{equation}
increase from zero with time, it remains smaller than $1$ for all times and inflation never ends.  Nevertheless, that is not our concern here and it is sufficient for our purposes that every mode of interest starts deep inside the horizon and exists the horizon at some time.  We can now calculate $\delta\nu^2$ from eq.~\eqref{dnu2-3/2} or directly from eq.~\eqref{nu2-test} and obtain
\begin{equation}
\delta\nu^{2} = \frac{-4c}{H_0\tau - c} \approx - \frac{4c}{H_0\tau} \left( 1 + \frac{c}{H_0\tau} \right).
\end{equation}
This expression, when plugged in eq.~\eqref{du-Green}, gives $\delta u$.  Although the integral can be performed analytically, the resulting expression is too long and not illuminating. Instead we have plotted $|u_k(\tau)|$ in figure~\ref{fig:green-method}, which shows good agreement with the numerical calculation. We have also used it to calculate $|u_{k_\sigma}(\tau)|^2$ and from that plot the noise power spectrum in figure~\ref{fig:variable-amplitude}.  Admittedly, the very simple approximation of eq.~\eqref{corr-Hksigma} works better than Green's method in this case. This is due to the growth of the source $S\propto\bar u$ in time (which enhances the perturbations and hence makes for larger error in Green's method), as well as the fact that $H/2\pi$ is evaluated at superhorizon $k=k_{\sigma}(\tau)$ in the noise power spectrum (which makes eq.~\eqref{corr-Hksigma} more accurate).  However, this method can be useful for other purposes, especially when the mode function $u_k$ is required at an intermediate time, rather than late time, as can happen, e.g., in loop integral calculations.

\section{Backreaction Effect of Non-Vacuum States}\label{app:backreaction}

In this appendix we calculate the energy density
\begin{equation}\label{rho-x}
\rho = \frac12 \langle \Psi | \dot{\hat \phi}^2 + \frac1{a^2} |\nabla{\hat \phi}|^2 + m^2 \hat\phi^2 | \Psi \rangle
\end{equation}
in a state of the form~\eqref{Psi-C2}, with specific results in the case that $C_2$ is given by eq.~\eqref{C2-exp}.  The aim of this calculation is to compare $\rho$ with the background energy density $\bar\rho = 3M_P^2H^2$ in dS space, to see if backreaction effects are significant.

In order to calculate $\rho$, we insert the Fourier transform~\eqref{mode-fxn} of the field operators in eq.~\eqref{rho-x} to obtain
\begin{equation}\label{rho-k}
\rho(\b x, \tau) = \frac{1}{2a^2} \int \frac{d^3k_1 d^3k_2}{(2\pi)^3} \left[ \langle {\hat \phi}'_{\b k_1} {\hat \phi}'_{\b k_2} \rangle + (-\b k_1 \cdot \b k_2 + m^2 a^2) \langle {\hat \phi}_{\b k_1} {\hat \phi}_{\b k_2} \rangle \right] e^{i(\b k_1 + \b k_2) \cdot \b x}.
\end{equation}
Given the form of $|\Psi\rangle$ in eq.~\eqref{Psi-C2}, there are three contributions to the matrix element $\langle {\hat \phi}_{\b k_1} {\hat \phi}_{\b k_2} \rangle$: one that is proportional to $|C_0|^2$, namely,
\begin{equation}\label{00}
\langle {\hat \phi}_{\b k_1} {\hat \phi}_{\b k_2} \rangle_{00} = |C_0|^2 |\phi_{k_1}|^2 \delta(\b k_1 + \b k_2),
\end{equation}
one that is proportional to $C_0^*C_2$ or $C_0C_2^*$, namely,
\begin{equation}\label{02}
\langle {\hat \phi}_{\b k_1} {\hat \phi}_{\b k_2} \rangle_{02} = 2 \Re \left[ C_0^* C_2(\b k_1, \b k_2) \phi_{k_1} \phi_{k_2} \right],
\end{equation}
and one that is proportional to $C_2^*C_2$, namely,
\begin{equation}\label{22}
\begin{aligned}
\langle {\hat \phi}_{\b k_1} {\hat \phi}_{\b k_2} \rangle_{22} &= \frac12 |\phi_{k_1}|^2 \delta(\b k_1 + \b k_2) \int |C_2(\b q_1, \b q_2)|^2 d^3q_1 d^3q_2 \\
&+ 2 \Re \left[ \phi_{k_1} \phi_{k_2}^* \int C_2(\b k_1, \b q) C_2^*(\b q, -\b k_2) d^3q \right].
\end{aligned}
\end{equation}
Eqs.~\eqref{02} and \eqref{22} are derived under the extra assumption $C_2(-\b k_1, -\b k_2) = C_2(\b k_1, \b k_2)$ (which is of course satisfied by the $C_2$ of eq.~\eqref{C2-exp}); otherwise they are general and valid in any FLRW background to all orders in $\varepsilon$.  There are similar expressions for $\langle {\hat \phi}'_{\b k_1} {\hat \phi}'_{\b k_2} \rangle$, too, with $\phi_k$s replaced by $\phi'_k$s.  It will be useful in the sequel to note that since the quantities inside $\Re$ in eqs.~\eqref{02} and \eqref{22} are unchanged under $\b k_{1,2} \to -\b k_{1,2}$, we can take this $\Re$ outside the integral in eq.~\eqref{rho-k}.

There are some parts of the three contributions above that we call `coherent' terms (those containing $|\phi_{k_1}|^2$ in which time oscillations are absent), and other parts that we call `incoherent' terms (those containing $\phi_{k_1} \phi_{k_2}$ or $\phi_{k_1} \phi_{k_2}^*$ which have time oscillations of the form $e^{-i(k_1+k_2)\tau}$ or $e^{-i(k_1-k_2)\tau}$).  The coherent terms appear in $\langle \ldots \rangle_{00}$ and $\langle \ldots \rangle_{22}$ and are proportional to $\delta(\b k_1+\b k_2)$, so we can easily perform one of the $k$-integrals in eq.~\eqref{rho-k} and obtain
\begin{equation}\label{rho-coherent}
\rho_{\rm coh} = \frac{1}{2a^2} \int \frac{d^3k}{(2\pi)^3} \left[ |\phi'_k|^2 + (k^2 + m^2 a^2) |\phi_k|^2 \right].
\end{equation}
At early times $\phi_k \to e^{-ik\tau}/a\sqrt{2k}$, $\phi'_k \to -ik \phi_k$, and the mass term is negligible.  The resulting integral has a quartic UV divergence like the usual zero-point energy contribution from $|0\rangle$.  Applying a physical cutoff at $k_{\rm max}/a = \Lambda$, it yields the standard result
\begin{equation}\label{rho-CC}
\rho_{\rm coh} = \frac{\Lambda^4}{16\pi^2},
\end{equation}
which should be absorbed in the background cosmological constant by an appropriate renormalization.

Let us now consider the incoherent contributions to $\rho$.  They appear in $\rho_{02}$ which arises from eq.~\eqref{02}, and also in $\rho_{22}$ which arises from eq.~\eqref{22} (but only its second line).  We show that the incoherent terms can be simplified when $C_2(\b q_1, \b q_2)$ depends on the magnitudes $q_{1,2}$ in a separable manner, i.e., $C_2(\b q_1, \b q_2) = f(q_1) f(q_2)$, as is the case, e.g., in eq.~\eqref{C2-exp} where
\begin{equation}\label{f-exp}
f(q) = \sqrt{\frac{\sqrt2 \varepsilon}{\pi Q^3}} e^{-q/Q}.
\end{equation}
Let us first consider $\rho_{02}$, which is purely incoherent.  Because of the separability of $C_2$, two of the three terms in eq.~\eqref{rho-k} can be written as a product of two integrals, each of the form
\begin{equation}
\int f(k)\phi_k e^{i \b k \cdot \b x} d^3k = 4\pi \int_0^\infty f(k)\phi_k \frac{\sin kx}{kx} k^2dk.
\end{equation}
The middle term in eq.~\eqref{rho-k} contains an extra factor $\b k_1 \cdot \b k_2$, which prevents factorization into two integrals.  But we can use the identity $\hat{\b k}_1 \cdot \hat{\b k}_2 = \cos\theta_1 \cos\theta_2 + \sin\theta_1 \sin\theta_2 \cos(\varphi_1-\varphi_2)$ (where $\theta_i, \varphi_i$ are the usual spherical coordinates of $\b k_i$, when $\hat{\b x}$ is chosen as the $z$-axis) and note that the $\varphi$-integrals vanish, so that again we have a product of two integrals, now of the form
\begin{equation}
\int f(k)\phi_k k\cos\theta e^{ikx\cos\theta} k^2dkd\cos\theta d\varphi = 4\pi i \int_0^\infty f(k)\phi_k \left( \frac{\sin kx}{k^2x^2} - \frac{\cos kx}{kx} \right) k^3dk.
\end{equation}
Combining these results, we find
\begin{equation}
\rho_{02} = 4\Re \left[ C_0^* ({\cal I}_1^2 + {\cal I}_2^2 + {\cal I}_3^2 ) \right],
\end{equation}
where
\begin{equation}\label{I1}
{\cal I}_1 = \frac{1}{a} \int_0^\infty f(k) \phi'_k(\tau) \frac{\sin kx}{kx} \frac{k^2dk}{\sqrt{2\pi}},
\end{equation}
\begin{equation}\label{I2}
{\cal I}_2 = \frac{1}{a} \int_0^\infty f(k) \phi_k(\tau) \left( \frac{\sin kx}{k^2x^2} - \frac{\cos kx}{kx} \right) \frac{k^3dk}{\sqrt{2\pi}},
\end{equation}
\begin{equation}\label{I3}
{\cal I}_3 = m \int_0^\infty f(k) \phi_k(\tau) \frac{\sin kx}{kx} \frac{k^2dk}{\sqrt{2\pi}}.
\end{equation}
The incoherent piece of $\rho_{22}$ (from the second line of eq.~\eqref{22}) is obtained in exactly the same manner:
\begin{equation}
\rho_{22}^{\rm inc} = 4 \Re \left[ F ({\cal I}_1^* {\cal I}_1 + {\cal I}_2^* {\cal I}_2 + {\cal I}_3^* {\cal I}_3) \right],
\end{equation}
where $F = \int |f(q)|^2 d^3q = \sqrt{2(1-|C_0|^2)}$ (which follows from the normalization of $C_2$ --- with the normalization given below eq.~\eqref{f-exp}, this yields $F = \sqrt2 \varepsilon$).  Thus the net value of the incoherent contributions to the energy density is
\begin{equation}\label{rho-incoherent}
\rho_{\rm inc} = 4\Re \left[ C_0^* ({\cal I}_1^2 + {\cal I}_2^2 + {\cal I}_3^2 ) \right] + 4 F ( |{\cal I}_1|^2 + |{\cal I}_2|^2 + |{\cal I}_3|^2 ).
\end{equation}

The total amount of the energy density is of course the sum of $\rho_{\rm coh}$ from eq.~\eqref{rho-coherent} and $\rho_{\rm inc}$ from eq.~\eqref{rho-incoherent}.  But at early times, the former becomes uniform (c.f.\ eq.~\eqref{rho-CC}) and gets absorbed in the background; thus it is only the latter that we should worry about when investigating backreaction effects.  It is possible to obtain closed form expressions for ${\cal I}_{1,2,3}$ in the case that $f$ is given by eq.~\eqref{f-exp} with the mode function of a massless field in dS space, but since the results are complicated and not illuminating we do not present them here.  Instead, we report their early time limits (${\cal I}_3=0$ in the massless case):
\begin{equation}\label{I-t>>x}
{\cal I}_1 \to \frac{15}{16} e^{i\pi/4} f(0) \sqrt{H^7 a^3}, \qquad {\cal I}_2 \to -\frac{3}{2} (aHx) {\cal I}_1.
\end{equation}
Thus we find that, in this case, and as $a\to0$:
\begin{equation}\label{rho-t>>x}
\rho_{\rm inc} \sim \varepsilon^2 Q^{-3} H^7 a^3 = \varepsilon^2 \sigma^{-3} H^4 \exp[3(N-N_Q)].
\end{equation}
Thus at times much earlier than $N_Q$, it seems that the backreaction on the energy density from the scalar field becomes irrelevant.  Notice how the distinction between coherent and incoherent terms played a crucial role here: While the former are time-independent, the latter damp as $\tau\to-\infty$ due to their rapidly oscillating integrands.

\begin{figure}
\centering
\includegraphics[width=0.6\textwidth]{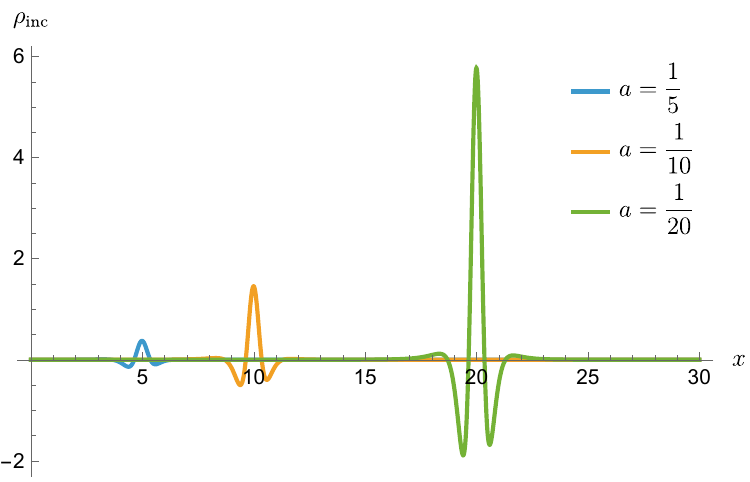}
\caption{Plot of $\rho_{\rm inc}(\tau,x)$ as a function of the radial distance $x$, at three times $a=\frac15, \frac1{10}, \frac1{20}$.  We have used the full eq.~\eqref{rho-incoherent} with eq.~\eqref{f-exp}, $\varepsilon=0.1$, $m=0$, $H=1$ and $Q=1$.} 
\label{fig:rho}
\end{figure}

The presence of $x$ in the limit of ${\cal I}_2$ in eq.~\eqref{I-t>>x} is an alarming sign that this approximation breaks at larger distances.  Indeed, the early time limits in eq.~\eqref{I-t>>x} are valid only when $|\tau|$ is much larger than any other length scale in the integrands, i.e., $|\tau| \gg x,Q^{-1}$.  We therefore need to plug in $x=-\tau$ first and then take the early time limit $\tau\to-\infty$, which yields:
\begin{equation}\label{I-t=x}
{\cal I}_1|_{x=-\tau} \to \frac{3}{16} f(0) \frac{H\sqrt{Q^5}}a, \qquad {\cal I}_2|_{x=-\tau} \to -{\cal I}_1|_{x=-\tau}.
\end{equation}
Inserting these values in eq.~\eqref{rho-incoherent}, we find, to leading order in $\varepsilon$, that at early times
\begin{equation}
\rho_{\rm inc}(x=-\tau) \sim \varepsilon \left( \frac{HQ}{a} \right)^2.
\end{equation}
In fact, if we numerically plot it, as we have done in figure~\ref{fig:rho}, we find that $\rho_{\rm inc}$ is maximal at $x=-\tau$ with a value proportional to $1/a^2$, as suggested by our calculation.  We can now compare this maximum energy density with the background value:
\begin{equation}
\frac{\rho_{\rm inc}(x=-\tau)}{\bar\rho} \sim \varepsilon \left( \frac{Q}{a M_p} \right)^2 = \varepsilon \left( \frac{\sigma H}{a_Q M_p} \right)^2.
\end{equation}
Comparing this ratio with 1, we find the criterion $a_Q \gg \sqrt\varepsilon \sigma H/M_P$.  Incidentally, this is the same as the condition we obtained at the end of section~\ref{sec:state} by a naive argument, except with an extra factor $\sqrt\varepsilon$.  This additional factor makes the duration of the inflationary period before $N_Q$ even larger than $\sim 15$ $e$-folds.  We thus conclude that the backreaction effect at early times is under control even with mild values of $\varepsilon$.

We now extend the above conclusion beyond the particular $f(q)$ of eq.~\eqref{f-exp}.  We assume, as usual, that the spacetime is asymptotically de~Sitter in the past (so that $\tau\to-1/a/H$ and $\phi_k \to e^{-ik\tau}/a\sqrt{2k}$).  Thus eqs.~\eqref{I1}--\eqref{I3} can be thought of as Fourier transforms from $k$-space to $\tau$-space.  Since the asymptotic behavior ($\tau\to-\infty$) of the Fourier transform is determined by the behavior of the integrand at its singularities and discontinuities, we only need to know what happens as $k\to0+$, as long as $f(k)$ is well-behaved elsewhere.  We also assume that $f(0)$ is finite.  Then a simple dimensional argument gives
\begin{equation}\label{I-t>>x-general}
{\cal I}_1 \sim \frac{1}{a^2} f(0) \tau^{-7/2}, \qquad {\cal I}_2 \sim \frac{x}{a^2} f(0) \tau^{-9/2}, \qquad {\cal I}_3 \sim \frac{m}{a} f(0) \tau^{-5/2}.
\end{equation}
With $a\to-1/H\tau$ asymptotically, we see that all of these vanish at early times, with the dominant behavior determined by ${\cal I}_1 \sim f(0) H^{7/2} a^{3/2}$.  Our previous result~\eqref{rho-t>>x} for the special case above is consistent with this general behavior.  Thus we find that for this class of states, $\rho \sim a^3$ and backreaction poses no problem at early times.  (There is no problem at late times either, since $\phi_k$ freezes and so all the three integrals in eqs.~\eqref{I1}--\eqref{I3} are under control.)

As before, the limits in eq.~\eqref{I-t>>x-general} are valid for $|\tau|\gg x$ and we should worry about larger values of $x$.  By inspecting the integrands in eqs.~\eqref{I1}--\eqref{I3}, it is evident that if $x=-\tau$, the oscillatory factor of $\phi_k(\tau)$ (i.e., $e^{-ik\tau}$) cancels the oscillatory factor of $\sin kx$ or $\cos kx$ (i.e., $e^{ikx}$).  The absence of oscillations prohibits damping at $\tau\to-\infty$ and  leads to an enhancement in the result of the integral.  It is now straightforward to compute 
\begin{equation}\label{I-t=x-general}
{\cal I}_1|_{x=-\tau} \to \frac{H\tilde Q}{4a}, \qquad {\cal I}_2|_{x=-\tau} \to -{\cal I}_1|_{x=-\tau}, \qquad {\cal I}_3|_{x=-\tau} \to \frac{i\tilde A mH}{4},
\end{equation}
where
\begin{equation}
\tilde A = \frac{1}{\sqrt\pi} \int_0^\infty f(k) k^{1/2} dk \qquad \text{and} \qquad \tilde Q = \frac{1}{\sqrt\pi} \int_0^\infty f(k) k^{3/2} dk.
\end{equation}
$\tilde A$ is dimensionless and $\tilde Q$ is a quantity with dimension of mass that corresponds to the characteristic scale of $f$ (for the $f$ of eq.~\eqref{f-exp}, we have $\tilde A = \sqrt{\varepsilon}/2^{3/4}\sqrt\pi$ and $\tilde Q = 3\sqrt{\varepsilon}Q/2^{7/4}\sqrt\pi$).  Thus the maximum energy density is given by
\begin{equation}
\rho_{\rm inc}(x=-\tau) \sim \left( \frac{H\tilde Q}{a} \right)^2,
\end{equation}
which is of the form we found earlier for a particular $f$.  Therefore, all of those conclusions carry over, with $Q$ replaced by $\tilde Q/\sqrt\varepsilon$, the characteristic scale of $f$.

We have thus confirmed that the backreaction effect is under control by a suitable choice of $\varepsilon$, and that even for $\varepsilon={\cal O}(1)$ with the fiducial values $H=10^{-5}M_P$ and $\sigma=10^{-2}$, we have room for about 15 $e$-folds of inflation.  This confirms the result of the naive analysis at the end of section~\ref{sec:state}, although these are two distinct criteria: one deals with comparing the energy density $\rho$ with $\bar\rho$, the other is a comparison of the typical energy $E$ of the particles with $M_P$.  Of course, the total energy computed from $E = \int \rho a^3 d^3x$ and eq.~\eqref{rho-k} is consistent with $E\sim Q/a$,\footnote{We should subtract $\rho_{\rm inc}$ which would give an infinite contribution to $E$ (corresponding to the product of the cosmological constant and the infinite volume of space).  The contribution from $\rho_{02}$ is subdominant at early times, and that of $\rho_{22}^{\rm inc}$ gives
\begin{equation}
E = \frac14 \int \frac{q_1+q_2}{a} |C_2(\b q_1, \b q_2)|^2 d^3q_1 d^3q_2.
\end{equation}
This is just the average energy of the states $|\b q_1, \b q_2 \rangle$ with their appropriate weights, which, for a $C_2$ peaked at $Q$, gives $E\sim Q/a$, as expected.  Note that we cannot write $\rho$ in an average form like this, because although $|\b q_1, \b q_2\rangle$ is an eigenstate of energy, it is not an eigenstate of energy density.} so there is nothing wrong with this estimate.  It is just that comparison of $E$ with $M_P$ is not the right criterion, although by a fortunate coincidence it gives the correct answer.


\bibliography{non-white}{}
\bibliographystyle{JHEP.bst}

\end{document}